\documentclass[aps,prd,floatfix,amsmath,amssymb,preprint,eqsecnum,nofootinbib]{revtex4}
\usepackage{graphicx}
\usepackage{dcolumn}
\usepackage{bm}
\usepackage{amsmath, amsfonts, amssymb}
\preprint{arXiv:1209.1637}

\newcommand{\beq}{\begin{equation}}
\newcommand{\eeq}{\end{equation}}
\newcommand{\ba}{\begin{array}{ccc}}
\newcommand{\ea}{\end{array}}
\newcommand{\nn}{\nonumber \\}

\def\bea{\begin{eqnarray}}
\def\eea{\end{eqnarray}}

\usepackage{setspace} 
\begin{document}
\title{Compressible quantum phases from\\ conformal field theories in 2+1 dimensions}
\author{Subir Sachdev}
\affiliation{Department of Physics, Harvard University, Cambridge MA
02138}

\date{September 13, 2012 \\
\vspace{1.6in}}
\begin{abstract}
Conformal field theories (CFTs) with a globally conserved U(1) charge $\mathcal{Q}$ can be deformed into
compressible phases by modifying their Hamiltonian, $\mathcal{H}$, by a chemical potential $\mathcal{H} \rightarrow
\mathcal{H} - \mu \mathcal{Q}$. We study 2+1 dimensional CFTs upon which an explicit $S$ duality mapping can be performed.
We find that this construction  leads naturally 
to compressible phases which are superfluids, solids,
or non-Fermi liquids which are more appropriately called `Bose metals' in the present context. 
The Bose metal preserves all symmetries and has Fermi surfaces of gauge-charged fermions, even
in cases where the parent CFT can be expressed solely by bosonic degrees of freedom.
Monopole operators are identified as order parameters of the solids, and the product of their magnetic charge and $\mathcal{Q}$ determines the area
of the unit cell.
We present implications for holographic theories on asymptotically 
AdS$_4$ spacetimes: $S$ duality and monopole/dyon fields play important roles in this connection.
\end{abstract}

\maketitle

\section{Introduction}
\label{sec:intro}

A powerful method of analyzing 
correlated systems of interest to condensed matter physics is the application of a chemical potential 
to ({\em i.e.\/} doping) conformal field theories in 2+1 dimensions (CFT3s) \cite{nernst}. This opens up a route to 
applying the advanced technology of the AdS/CFT correspondence \cite{Maldacena,GKP,Wittenads}. 
In the latter approach, the charge density conjugate to the
chemical potential is equated to an electric flux emanating from the boundary of a spacetime which is asymptotically AdS$_4$.
A central question in then the nature of the stable ground state in the presence of a boundary electric flux.

Gubser \cite{gubser} pointed out that the ground state of a doped CFT 
could be a superfluid. He described the condensation of a bulk
charged scalar in the presence of a AdS-Reissner-N\"ordtrom black brane solution with a charged horizon \cite{myers}. 
It is important to note, however, that a 
conventional superfluid does not have low energy excitations associated with such a horizon above its ground state.
Such a superfluid only appears when the infra-red (IR) geometry is confining, there is no horizon, and the boundary electric flux
is fully neutralized by the condensate \cite{tadashi,gary}.

A second class of compressible ground states of doped CFTs, known variously as `non-Fermi liquids', `strange metals' or `Bose metals', 
were proposed in \cite{liza} using arguments mainly from the CFT side. These states have Fermi surfaces of fermions carrying both the global
U(1) charge of the doped CFT, and the charges of deconfined gauge fields (in the condensed matter context, the latter gauge fields are invariably
`emergent').  Because of the gauge charges, the single-fermion Green's function is not a gauge-invariant observable, and so the Fermi surface
is partially ``hidden'' \cite{hyper}.
In the holographic context, these gauge-charged Fermi surfaces are possibly linked to electric
flux that goes past the horizon in the zero temperature limit \cite{tomjoe,ssffl,liza,ssfl,arcmp,hartnollrev,lizasean,iqballutt,mcgreevy,seanrad}. 
Particular holographic duals for these states \cite{kiritsis,trivedi,tadashi1,hyper},
were supported by evidence which matched the entropy density, numerous features of the entanglement entropy, and an inequality
on the critical exponents \cite{tadashi1,hyper}. 

Finally, from a condensed matter perspective, a natural ground state of a doped CFT is a solid (or a `crystal' or a `striped state'), in which
the doped charges localize in a regular periodic arrangement. Spatially modulated states have been found in the context 
of the AdS/CFT correspondence \cite{ooguri1,ooguri2,jerome1,jerome2,jerome3,iizuka} but in situations with parity violating terms or in the presence
of magnetic fields. We will see here that a version of such instabilities, after a $S$ dual mapping of CFTs in 2+1 dimensions \cite{witten,m2cft},
will yield solid phases of parity preserving CFTs in a chemical potential. The solid will choose its periodic density modulation 
so that there are an integer number
of doped charges per unit cell.

An important motivation for the present work was provided by the recent analysis by Faulkner and Iqbal \cite{fi}. 
They examined holographic duals of finite density quantum systems in 1+1 dimensions, and showed that monopole tunneling events
in the bulk led to oscillatory density-density correlations on the boundary. They identified these oscillatory correlations as the Friedel oscillations
of an underlying Fermi surface. Similar oscillations also appeared in deconfined phases of gauge theories coupled to 
fermionic matter at non-zero density \cite{rajesh}.
However, it should be noted that in one spatial dimension such oscillatory correlations are present also in superfluids and solids,
neither of which breaks any symmetry.

We are interested here in examining the role of monopoles on CFT3s in 2+1 dimensions,
and in the corresponding doped CFT3s. The monopole and dyon operators of such 
CFT3s are closely linked to their properties under $S$ duality transformations \cite{murthy,kapustin1,kapustin2,kapustin3,witten,hermelemono,hermeleo4,benna,willett}.
We will therefore present a reasonably complete description of two CFTs with global U(1) symmetries upon which the $S$ 
duality transformation can be explicitly carried out. In the absence of supersymmetry, such explicit transformations are
only possible in theories with abelian gauge fields, abelian global symmetries, and bosonic matter; our CFTs are two of the simplest
examples with such restrictions: the $XY$ model and the abelian $\mathbb{CP}^{N-1}$ model. 

After describing these CFT3s, we will dope them into compressible states 
by applying a chemical potential. In both cases, we easily find that such CFT3s can exhibit superfluid and solid phases.
Naturally, the superfluid phases break the global U(1) symmetry.  
We will show that the monopole operators serve as order parameters for the solid phases: condensation of monopoles
implies broken translational symmetry; this is similar to phenomena in insulating phases \cite{rsl,rsb,senthil}.
Furthermore, the magnetic charge of the monopole condensate will determine the size of the unit cell so that there are
an integer number of doped electric charges per unit cell.

However, our primary interest is in phases of doped CFTs which do not break any symmetries. Such phases appeared
in the previous analysis of CFTs with fermionic degrees of freedom \cite{liza} as non-Fermi liquid states with Fermi surfaces
of gauge-charged fermions. We will show here that essentially identical compressible 
phases also appear upon doping CFT3s whose local Lagrangian contains
only bosonic degrees of freedom. These compressible phases also contain gauge-charged Fermi surfaces of emergent fermionic degrees of
freedom. The advantage of our present bosonic starting point is that it will shed new light on the role of monopoles, dyons,
and $S$ duality on such phases, and this information is surely crucial in setting up a complete holographic theory. 
Given our bosonic formulation, we will call these non-Fermi liquid states `Bose metals'. This appellation is also apropos given
the similarity of our analysis to the Bose metal phases of lattice spin and boson models \cite{motfish,motfish2,motfish3,motfish4}.

\begin{figure}[h]
\begin{center}
 \includegraphics[width=5.5in]{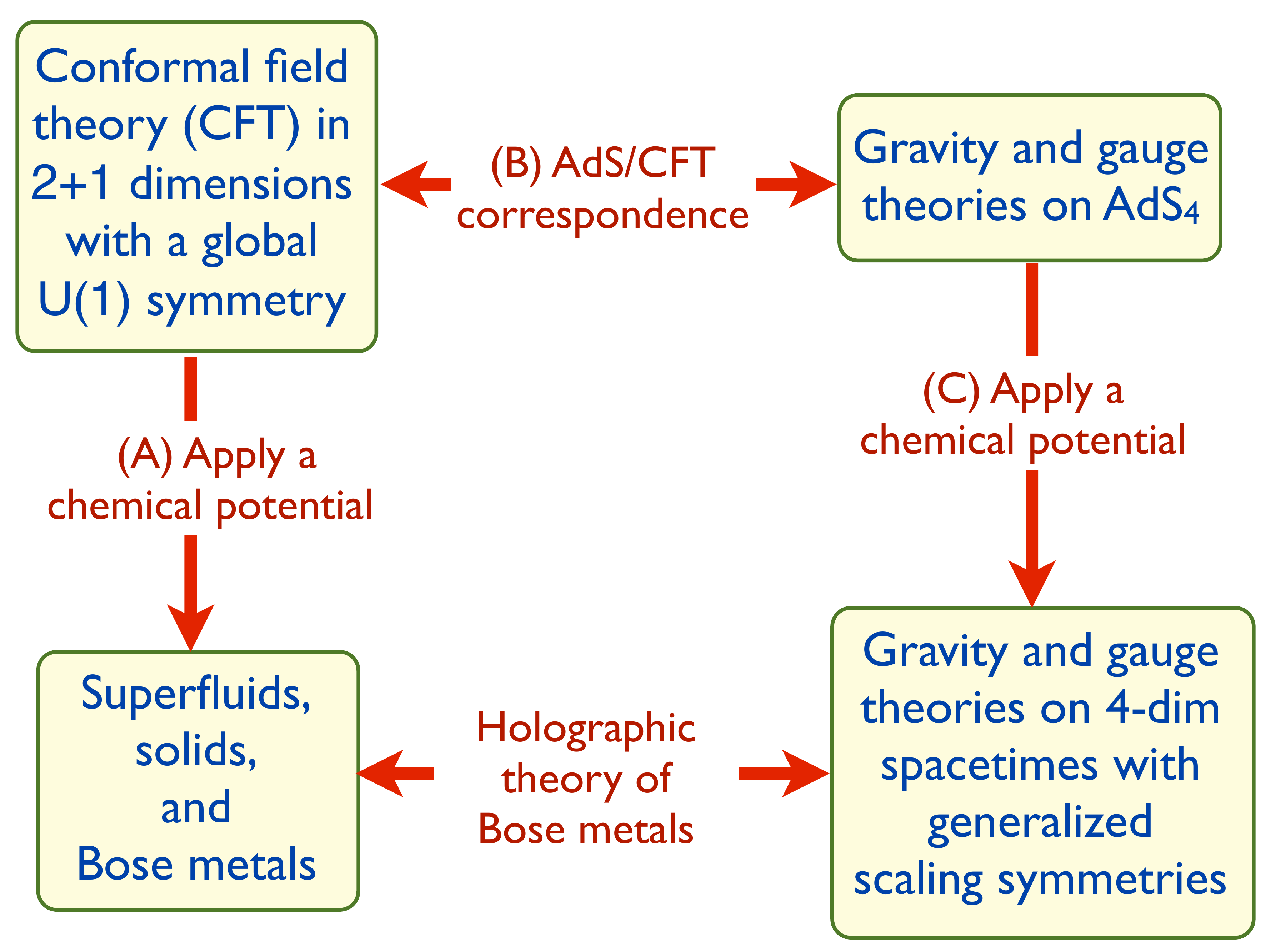}
 \caption{Connections between CFT3s and holography}
\label{fig:flowchart}
\end{center}
\end{figure}

It is useful to summarize the relationships between ideas discussed here in the flowchart in Fig.~\ref{fig:flowchart}.
We have so far discussed step A in Fig.~\ref{fig:flowchart}.
Armed with this understanding of the $S$ duality of CFT3s, and the possible phases of doped CFT3s, we move onto
holographic considerations. For the case of CFT3s, as in step B of Fig.~\ref{fig:flowchart},  
we will discuss features of the holographic theory on  AdS$_4$,
building on ideas of Witten \cite{witten} on the role of $S$ duality. 
We propose a bulk theory with fields corresponding to the electric, magnetic, and dyon operators
of the CFT3, and these couple to 3+1 dimensional U(1) Maxwell fields and their electromagnetic duals.
We will check our bulk theory by a comparison of its predictions for 3-point correlators with those of 
boundary CFT3s in Appendix~\ref{app:3pt}.
Then we apply a chemical potential, as in step C, by fixing the value of a Maxwell vector potential at the boundary of AdS$_4$.
We will discuss aspects of the resulting holographic theory, and note connections to the
phases obtained from a direct analysis of the doped CFT3s. 

We note that 
we will restrict our attention to models which preserve time-reversal and parity 
symmetries. So external magnetic fields coupling to the global U(1) charge
will not be allowed. Although after $S$ duality transformations some of our holographic solutions will contain background ``magnetic" fields, there is always a frame in which background is purely electric, and parity and time-reversal are preserved. 
We do not consider situations in which there is both a chemical potential and a magnetic field, leading to possible quantum Hall states.

We will begin by describing two model CFT3s with global U(1) symmetries 
in Section~\ref{sec:cft}: the $XY$ model and the abelian $\mathbb{CP}^{N-1}$ model. We will describe some of their properties, including identification
of their primary operators with electric, magnetic, and dyonic charges:
this will be important for the holographic formulation. 
We will apply a chemical potential to these CFT3s in Section~\ref{sec:dope}, as in step A of Fig.~\ref{fig:flowchart}. 
This will allow us to identify classes of phases which are possible in such situations. 
Finally, in Section~\ref{sec:ads}, we will discuss feature of the 
holographic realizations of these CFT3s (step B) and their compressible descendants (step C).

We will follow the convention of using indices $\mu,\nu \ldots$ for the 3 spacetime components, $a,b\ldots$ for the 4 directions of AdS$_4$,
and $i,j \ldots$ for the 2 spatial components. We use the Euclidean time signature throughout.

\section{Conformal field theories}
\label{sec:cft}

\subsection{$XY$ model}
\label{sec:xy}

We begin with the simplest possible interacting CFT3 (in 2+1 dimensions) with a conserved U(1) charge: the $XY$ model, described by the
Wilson-Fisher fixed point of the $\phi^4$ field theory
\beq
\mathcal{L}_{XY} = |\partial_\mu \phi|^2 + s |\phi|^2 + u |\phi|^4 .\label{lxy}
\eeq
This fixed point has one relevant operator, and we assume that either $s$ or $u$ has been tuned to place the field theory
at the conformally invariant point.
The complex field $\phi$ will serve as our superfluid order parameter. 
We can also consider it as an operator carrying unit `electric' charge, and we define
its correlator as
\beq
G_e (y) = \left \langle \phi^\ast (y) \phi (0) \right\rangle_{\mathcal{L}_{XY}}
\label{exy}
\eeq
This correlation function has a power-law decay for the CFT3, and this defines the scaling dimension of the electric operator.
The CFT3 has a conserved charge $\mathcal{Q}$ associated with the current
\beq
J_\mu = i \left( \phi^\ast \partial_\mu \phi - \phi \partial_\mu \phi^\ast \right) \label{defjmu}
\eeq
where 
\beq
\mathcal{Q} = \int d^2 x J_\tau
\eeq
and $\mathcal{Q} = 1$ is the electric charge of quanta of $\phi$.
This current has the correlator (after subtracting a contact term)
\beq
\left\langle J_\mu (k) J_\nu (-k) \right\rangle_{\mathcal{L}_{XY}} = - \frac{1}{g^2}\,  |k| \left( \delta_{\mu\nu} - \frac{k_\mu k_\nu}{k^2} \right), \label{JJ}
\eeq
where $g$ is a universal number characteristic of the $XY$ CFT3, and $k_\mu$ is a 3-momentum.

We can also define a monopole operator for this CFT3 as follows. First, couple $\phi$ to a background gauge field $\alpha_\mu$
\beq
\mathcal{L}_{XY} [\alpha] = |(\partial_\mu - i  \alpha_\mu) \phi|^2 + s |\phi|^2 + u |\phi|^4 \label{lxyalpha}
\eeq
where we follow the convention of indicating source/background fields which are not integrated over as arguments of the Lagrangian.
We choose the magnetic field 
\beq
\beta_\mu = \epsilon_{\mu\nu\lambda} \partial_\nu \alpha_\lambda \label{betaalpha}
\eeq
to be 
sourced by monopoles of magnetic charge $\widetilde{\mathcal{Q}} = \pm 2 \pi$ at $x=0$ and $x=y$
\beq
\partial_\mu \beta_\mu = 2\pi  \, \delta^3 (x) - 2\pi \, \delta^3 (x-y). \label{divbeta}
\eeq
The monopole charge of $2 \pi$ is required by the Dirac quantization condition. 
Note that the monopole so defined has a subtle difference from those considered earlier \cite{murthy,hermelemono,hermeleo4},
which were monopoles in a dynamical gauge flux. Here we have a {\em background\/} gauge flux, coupling to the field theory by gauging
a global symmetry. 
Such monopole background fields were
also discussed recently by Kapustin and Willett \cite{willett}.
The monopole correlation function is
\beq
G_m (y) = \frac{ \int \mathcal{D} \phi \exp \left( - \int d^3 x \mathcal{L}_{XY} [\alpha] \right)}{\int \mathcal{D} \phi \exp \left( - \int d^3 x \mathcal{L}_{XY}  \right)} \label{mxya}
\eeq
This correlation function has a power-law decay for the CFT3 \cite{murthy}, and this defines the 
scaling dimension of the monopole operator .
This monopole scaling dimension was computed in \cite{murthy,hermelemono} in the large $N$ limit of a theory with $N$ copies
of the $\phi$ field; as we just noted, these computations also include a fluctuating gauge field, but this plays no role in the leading large $N$ limit.

We can also consider multi-point correlators of the operators defined above. A convenient probe of the structure of the theory
is provided by 3-point correlators of the current $J_\mu$ with the matter fields. Thus we have the correlator with the electric operators
\beq
\left\langle J_\mu (p) \phi^\ast (k_1) \phi (k_2) \right\rangle_{\mathcal{L}_{XY}} \label{xy3a}
\eeq
(the arguments of the fields are momenta),
and also with the monopole operators
\beq
 \frac{ \int \mathcal{D} \phi \, J_\mu (w) \exp \left( - \int d^3 x \mathcal{L}_{XY} [\alpha] \right)}{\int \mathcal{D} \phi \exp \left( - \int d^3 x \mathcal{L}_{XY}  \right)} \label{xy3b}
\eeq
(the arguments of the fields are spacetime co-ordinates).
Both these correlators are computed in Appendix~\ref{app:3pt}, which also provides a comparison with results from holography.

It is interesting to determine the conformal transformations of the monopole operator as defined above.
For this, we need to specify the background field $\beta_\mu$ more completely. Being a ``magnetic'' field, 
it is natural to have $\beta_\mu$ transform
as a vector of scaling dimension 2 (more properly, it is a 2-form field); 
it can then be checked that the divergence condition in (\ref{divbeta}) is indeed conformally invariant.
We also need zero curl conditions: unlike (\ref{divbeta}), the conformally invariant form of these conditions depends upon the
spacetime metric. With the conformally flat metric $ds^2 = \Omega^{-2} (x) d x_\mu dx_\mu$, the zero curl conditions are
\beq
\epsilon_{\mu\nu\lambda} \partial_\nu ( \Omega(x) \beta_\nu (x)) = 0. \label{curl}
\eeq
The equations (\ref{curl}) and (\ref{divbeta}) define $\beta_\mu$ in
a conformally invariant manner. In this way we obtain a definition
of the monopole operator which transforms like an ordinary scalar under all conformal transformations. Note that our definition is intrinsically
non-local as it involves a determination of a non-zero $\beta_\mu (x)$ at all spacetime points; 
the issue of the non-locality of the monopole operator will appear again when we discuss holography in Section~\ref{sec:ads}.

\subsubsection{$S$ duality}
\label{sec:sxy}

As reviewed in Appendix~\ref{app:xy}, application of $S$ duality to the $XY$ model yields the abelian Higgs model \cite{peskin,dasgupta}
\beq
\mathcal{L}_{XY}^S = |(\partial_\mu - i a_\mu) \psi|^2 + s |\psi|^2 + u |\psi|^4 + \frac{1}{2e^2} \left( \epsilon_{\mu\nu\lambda} \partial_\nu a_\lambda \right)^2 \label{xyS}
\eeq
which provides an alternative description of the same CFT3. The conserved U(1) current in (\ref{defjmu}) can now be written as
\beq
J_\mu =  \frac{i}{2 \pi} \epsilon_{\mu\nu\lambda} \partial_\nu a_\lambda .
\label{defjmu2}
\eeq
The factor of $i$ is a consequence of working in the Euclidean signature, and the exchange of electric and magnetic degrees of freedom under $S$ duality.
A gauge-invariant two-point correlator of the field $\psi$  
yields the same $G_m$ as in (\ref{mxya}), as shown in Appendix~\ref{app:xy}:
\beq
G_m (y) = \left\langle \psi^\ast (y) \exp \left( -\frac{i}{2 \pi} \int d^3 x \, a_\mu \, \beta_\mu \right) \psi (0) \right\rangle_{\mathcal{L}_{XY}^S} ;
\label{mxyb}
\eeq
it is easy to verify that the above correlation function is gauge invariant after using (\ref{divbeta}). 
From this correlator, we can identify the gauge-invariant monopole operator
as 
\beq
\psi \, \mathcal{M}_a \quad;\quad \widetilde{\mathcal{Q}}=2 \pi, \label{psima}
\eeq
where the operator 
$\mathcal{M}_a$ is defined from (\ref{mxyb}) as an insertion which couples the gauge field $a_\mu$ to the
magnetic flux of a monopole via a Chern-Simons term; such an insertion also appeared in the analysis of supersymmetric CFT3s by Kapustin and
Willett \cite{willett}. Explicitly we have the various representations
\bea
\mathcal{M}_a (y) \mathcal{M}_a^\dagger (0) &=& \exp \left(  \frac{i}{2 \pi} \int d^3 x \, a_\mu (x) \epsilon_{\mu\nu\lambda} \partial_\nu \alpha_\lambda (x)  \right) \nn
&=& \exp \left(  \frac{i}{2 \pi} \int d^3 x \, a_\mu (x) \beta_\mu (x)  \right); \nn
\mathcal{M}_a (y) &=& \exp \left(  \frac{i}{2 \pi} \int d^3 x \frac{a_\mu (x) (x_\mu - y_\mu)}{2 |x - y|^3} \right). \label{ma}
\eea
Note the non-local structure in the definition.
In the Chern-Simons formulation, the boundary term obtained after the gauge transformation of the Chern-Simons term reduces to 
contributions on the surfaces of small spheres surrounding $x=y$ and $x=0$, and this cancels the gauge transformation of $\psi (y) \psi^\ast (0)$.
See also similar constructions in \cite{dirac,sondhi}.

It is useful to interpret the operator (\ref{psima}) by the state-operator correspondence of CFT3s \cite{hermelemono}. For this, we radially quantize the CFT3
on surface of a sphere $S^2$. Then we see that $\psi$ creates a single quantum carrying a unit $a_\mu$ electric charge, delocalized over the spherical surface.
While $\mathcal{M}_a$ is a background charge density, with a term $i \int (d \Omega/(4 \pi)) a_\tau$ (where $d \Omega$ is the spherical solid angle);
the integral over $a_\tau$ merely projects the Hilbert space to states with unit electric charge.
Thus from the point of view of the scalar QED theory in (\ref{xyS}), $\psi \, \mathcal{M}_a $ is a gauge
invariant ``electric'' operator, whereas it is a magnetic operator from the perspective of the direct $XY$ model; our notation will always reflect the perspective
of the direct theory.
In previous work \cite{rantner,jinwu}, Wilson line operators have been used to obtain gauge-invariant correlators of matter
fields like $\psi$, but these are path-dependent, don't have simple conformal transformation properties,
don't define point-like operators which can be used in the state-operator correspondence, and don't appear in our duality analysis.

Finally, to close the circle of dualities, we have a representation of the electric correlator in (\ref{exy}) in terms of a monopole background
for $\mathcal{L}_{XY}^S$ \cite{peskin}, as described in Appendix~\ref{app:xy}:
\beq
G_e (y) = \frac{ \int \mathcal{D} \psi \mathcal{D} a_\mu \exp \left( - \int d^3 x \mathcal{L}_{XY}^S [\alpha] \right)}{ \int \mathcal{D} \psi \mathcal{D} a_\mu \exp \left( - \int d^3 x \mathcal{L}_{XY}^S  \right)}, \label{exyb}
\eeq
where
\beq
\mathcal{L}_{XY}^S [\alpha] = |(\partial_\mu - i a_\mu - i  \alpha_\mu) \psi|^2 + s |\psi|^2 + u |\psi|^4 . \label{lsxyalpha}
\eeq
This is actually a traditional local ``monopole'' correlator for the scalar QED in (\ref{xyS}), which maps to the electric correlator of (\ref{lxy}).
The associated scaling dimension was computed using the above monopole insertion method 
in \cite{murthy,hermelemono} in the large $N$ limit of a theory with $N$ copies
of the $\psi$ field; the fluctuating gauge field $a_\mu$ was included in these computations, but this will modify the scaling dimension
only at order $1/N$.

Similar $S$ duality 
mappings also apply to the 3-point correlators of the current $J_\mu$ and the matter fields, such as those in (\ref{xy3a}) and (\ref{xy3b}), and are 
discussed in Appendix~\ref{app:3pt}.

Let us also note that we can define a U(1) current associated with the $\psi$ matter field
\beq
\widetilde{J}_\mu = 2 \pi i \left( \psi^\ast (\partial_\mu -i a_\mu) \psi - \psi (\partial_\mu + i a_\mu) \psi^\ast \right) \label{deftjmu}
\eeq
In the scaling limit, where $e^2 \rightarrow \infty$ in (\ref{xyS}), the equation of motion of $a_\mu$ imposes the constraint $\widetilde{J}_\mu = 0$. 
Clearly, this implies that the two-point correlator of $\widetilde{J}_\mu$ must also vanish. It is then easy to show diagrammatically
that the {\em irreducible}  $\widetilde{J}_\mu$ correlator (with respect to a $a_\mu$ propagator) 
is non-zero, and equal to the inverse \cite{m2cft,mpaf2,wenzee} of the correlator of $J_\mu$ in (\ref{defjmu2})
\beq
\left. \left\langle \widetilde{J}_\mu (k) \widetilde{J}_\nu (-k) \right\rangle_{\mathcal{L}_{XY}^S}  \right|_{irr} = 
- {g^2}\,  |k| \left( \delta_{\mu\nu} - \frac{k_\mu k_\nu}{k^2} \right). \label{tJJ}
\eeq

To summarize, the $XY$ CFT3 has a global U(1) symmetry, an electric operator $\phi$,
and a magnetic operator $\psi \mathcal{M}_a$ (as we noted, this terminology reflects the perspective of 
the $XY$ model in (\ref{lxy}), and not its $S$ dual in (\ref{xyS})). 
Gauge-invariant correlators for these two operators can be written in both the direct
and $S$ dual formulations of the field theory. Such operators also obey nontrivial monodromy properties \cite{max1}, 
but we will not
explore this aspect here.

\subsection{Abelian $\mathbb{CP}^{N-1}$ model}
\label{sec:cp}

The $XY$ CFT3 will not be rich enough to display all the possible phases of CFT3s in the presence of a chemical potential.
So we consider 
the `easy-plane' or abelian $\mathbb{CP}^{N-1}$ model for which explicit $S$ duality transformations can be performed 
(without supersymmetry) \cite{mv,balents}.
This is also the simplest CFT3 which is both a gauge theory and has global U(1) symmetries.
It should be noted that the existence of a conformally-invariant fixed point for this specific 
field theory has not been conclusively demonstrated for the simplest $N=2$ case \cite{mv2}.
However, it is clear that such a CFT3 does exist in a large $N$ limit \cite{mv,senthil}, and
all of our analysis can be extended to general $N$ \cite{balents}. 
However, in the interests of simplicity, we will restrict ourselves to the
simplest $\mathbb{CP}^1$ case, and work with the assumption that this CFT3 does exist.

In the direct formulation, the degrees of freedom are two complex scalars, $z_1$ and $z_2$, and a non-compact U(1) gauge field $b_\mu$ with Lagrangian
\bea
\mathcal{L}_{\mathbb{CP}} &=& \left|\left(\partial_\mu - i b_\mu\right) z_1 \right|^2 +
\left|\left(\partial_\mu - i b_\mu\right) z_2 \right|^2  + \frac{1}{2e^2} \left( \epsilon_{\mu\nu\lambda} \partial_\nu b_\lambda \right)^2
+ \mathcal{L}_{z,{\rm loc}} \nn
\mathcal{L}_{z,{\rm loc}} &=& s \left(
|z_1|^2 + |z_2 |^2 \right) + u
\left(|z_1|^2 + |z_2|^2 \right)^2 + v |z_1|^2 |z_2 |^2  \label{lcp}
\eea
with $u>0$ and $-4u<v<0$. For these negative values of $v$, the
phase for $s$ sufficiently negative has $|\langle z_1 \rangle | =
|\langle z_2 \rangle | \neq 0$. We assume that the one relevant perturbation at the critical point has been tuned to obtain a CFT3.

This theory actually has two global U(1) symmetries, and associated conserved currents.
The first is the ordinary global symmetry
\beq
\mathcal{Q}_1: \quad z_1 \rightarrow z_1 e^{i \theta} \quad, \quad z_2 \rightarrow z_2 e^{-i \theta}
\eeq
The conserved current is 
\beq
J_{1\mu} = i \left( z_1^\ast (\partial_\mu - i b_\mu) z_1 - z_1
(\partial_\mu + i b_\mu) z_1^\ast \right) - i \left( z_2^\ast
(\partial_\mu - i b_\mu) z_2 - z_2 (\partial_\mu + i b_\mu) z_2^\ast
\right), \label{j1z}
\eeq
and
\beq
\mathcal{Q}_1 = \int d^2 x J_{1 \tau}.
\eeq
However, there is also a conserved `topological' U(1) current 
\beq
J_{2\mu} = \frac{i}{\pi} \epsilon_{\mu\nu\lambda} \partial_\nu b_\lambda \label{j2z}
\eeq
and a corresponding topological charge
\beq
\mathcal{Q}_2 = \int d^2 x J_{2\tau}
\eeq
Another significant symmetry of $\mathcal{L}_{\mathbb{CP}}$ is the $\mathbb{Z}_2$ symmetry under which $z_1 \leftrightarrow z_2$.
Note that $J_{1 \mu}$ is odd under this symmetry, while $J_{2\mu}$ is even. This prohibits a bilinear coupling between these current
operators.

As reviewed in Appendix~\ref{app:cp1}, we can perform the $S$ dual mapping on {\em both} U(1) symmetries.
This yields the a theory with the same Lagrangian (unlike the situation for the XY model), but now expressed in terms of 
complex scalars $w_1$ and $w_2$, and a U(1) gauge field $a_\mu$, which will have different physical interpretations.
So the
abelian $\mathbb{CP}^1$ model is self-dual \cite{mv} and has the Lagrangian
\beq
\mathcal{L}_{\mathbb{CP}}^S = \left|\left(\partial_\mu - i a_\mu\right) w_1 \right|^2 +
\left|\left(\partial_\mu - i a_\mu\right) w_2 \right|^2  + \frac{1}{2e^2} \left( \epsilon_{\mu\nu\lambda} \partial_\nu a_\lambda \right)^2
+ \mathcal{L}_{w,{\rm loc}}. \label{lcps}
\eeq
The values of the non-universal couplings in $\mathcal{L}_{w,{\rm loc}}$ have been modified from before.
 The key feature of this dual representation is that the roles
of the global and topological U(1) currents have been interchanged. Thus the currents in (\ref{j1z}) and (\ref{j2z}) now
have the representation
\beq
J_{2\mu} = i \left( w_1^\ast (\partial_\mu - i a_\mu) w_1 - w_1
(\partial_\mu + i a_\mu) w_1^\ast \right) - i \left( w_2^\ast
(\partial_\mu - i a_\mu) w_2 - w_2 (\partial_\mu + i a_\mu) w_2^\ast
\right),
\eeq
and
\beq
J_{1\mu} = \frac{i}{\pi} \epsilon_{\mu\nu\lambda} \partial_\nu a_\lambda. \label{J1a}
\eeq

Let us now turn to an identification of the electric and magnetic operators associated with these  U(1) symmetries. 
By analogy with the $XY$ model, the simplest choices for electric operators are gauge invariant combinations of the matter
fields which carry global charges
\bea
 z_2^\ast z_1 \quad&;&\quad \mathcal{Q}_1 = 2, \, \mathcal{Q}_2 = 0 \nn
 w_2^\ast w_1 \quad&;&\quad \mathcal{Q}_1 = 0, \, \mathcal{Q}_2 = 2 .\label{phi12}
\eea 
Similarly, following the analysis for the $XY$ model, we can write down gauge-invariant magnetic operators
which are electrically neutral but carry magnetic charges (see Appendix~\ref{app:cp1})
\bea
w_1 w_2 \, \mathcal{M}_a^2 \quad&;&\quad \widetilde{\mathcal{Q}}_1 = 2 \pi, \, \widetilde{\mathcal{Q}}_2 = 0 \nn
 z_1 z_2 \, \mathcal{M}_b^2 \quad&;&\quad \widetilde{\mathcal{Q}}_1 = 0, \, \widetilde{\mathcal{Q}}_2 = 2 \pi .\label{psimab}
\eea 
However, examination of the operators in (\ref{phi12}) and (\ref{psimab}) shows that all of these can be written as 
composites of simpler {\em dyonic \/} operators that carry electrical charges of one U(1) symmetry and magnetic charges of the other U(1) symmetry {\em i.e.\/} the operator product expansion of pairs of dyonic operators will produce operators in
(\ref{phi12}) and (\ref{psimab}).
We denote the magnetic charges by $\widetilde{\mathcal{Q}}_1$ and $\widetilde{\mathcal{Q}}_2$, and then the primary
dyonic operators are
\bea
 z_1 \, \mathcal{M}_b \quad&;&\quad \mathcal{Q}_1 = 1, \, \mathcal{Q}_2 = 0,  \, \widetilde{\mathcal{Q}}_1 = 0, \, \widetilde{\mathcal{Q}}_2 =  \pi
 \nn
 z_2 \, \mathcal{M}_b \quad&;&\quad \mathcal{Q}_1 = -1, \, \mathcal{Q}_2 = 0,  \, \widetilde{\mathcal{Q}}_1 = 0, \, \widetilde{\mathcal{Q}}_2 =  \pi \nn
  w_1 \, \mathcal{M}_a \quad&;&\quad \mathcal{Q}_1 = 0, \, \mathcal{Q}_2 = 1,  \, \widetilde{\mathcal{Q}}_1 =  \pi, \, \widetilde{\mathcal{Q}}_2 = 0 \nn
  w_2 \, \mathcal{M}_a \quad&;&\quad \mathcal{Q}_1 = 0, \, \mathcal{Q}_2 = -1,  \, \widetilde{\mathcal{Q}}_1 =  \pi, \, \widetilde{\mathcal{Q}}_2 = 0. \label{dyons}
\eea
We emphasize that these operators are all gauge-invariant.
Expressions for the correlators of all four operators can be obtained in both the direct and $S$ dual representations. 
As an example, let us consider the two-point correlator of $z_1 \mathcal{M}_b$: from its definition we have
as in (\ref{mxyb})
\beq
G_{z_1} (y) = \left\langle z_1^\ast (y) \exp \left( -\frac{i}{2 \pi} \int d^3 x \, b_\mu \, \beta_\mu \right) z_1 (0) \right\rangle_{\mathcal{L}_{\mathbb{CP}}} .
\label{z1a}
\eeq
We perform the $S$ duality mapping of this correlator in Appendix~\ref{app:cp1} and find
\beq
G_{z_1} (y) = \frac{ \int \mathcal{D} w_1 \mathcal{D} w_2 \mathcal{D} a_\mu \exp \left( - \int d^3 x \mathcal{L}_{\mathbb{CP}}^S [\alpha] \right)}{\int \mathcal{D} w_1 \mathcal{D} w_2 \mathcal{D} a_\mu \exp \left( - \int d^3 x \mathcal{L}_{\mathbb{CP}}^S  \right)}, \label{z1b}
\eeq
where a monopole background has been minimally coupled to $w_1$ via
\beq
\mathcal{L}_{\mathbb{CP}}^S [\alpha] = \left|\left(\partial_\mu - i a_\mu - i \alpha_\mu  \right) w_1 \right|^2 +
\left|\left(\partial_\mu - i a_\mu  \right) w_2 \right|^2  + \frac{1}{2e^2} \left( \epsilon_{\mu\nu\lambda} \partial_\nu a_\lambda \right)^2
+ \mathcal{L}_{w,{\rm loc}}. \label{lcp1}
\eeq
Similar expressions can be obtained for the remaining operators in (\ref{dyons}).

\section{Doped conformal field theories}
\label{sec:dope}

This section will describe step A of Fig.~\ref{fig:flowchart}, applied to the CFT3s described in Section~\ref{sec:cft}.

\subsection{$XY$ model}
\label{sec:dxy}

We apply a chemical potential, $\mu$, to the $XY$ model of (\ref{lxy})
\beq
\mathcal{L}_{XY}[\mu] = [(\partial_\tau + \mu) \phi^\ast ] [(\partial_\tau - \mu) \phi] + |\partial_i \phi|^2 + s |\phi|^2 + u |\phi|^4 
\label{dlxy}
\eeq
In the $S$ dual formulation of (\ref{dlxy}), this chemical potential also couples to $J_\tau$:
\beq
\mathcal{L}_{XY}^S [\mu]  = \mathcal{L}_{XY}^S
- \frac{\mu}{2\pi} \epsilon_{ij} \partial_i a_j . \eeq

The following subsections describe the superfluid and solid phases that can appear in such a CFT3 at a non-zero $\mu$.
A Bose metal phase is not possible in such a CFT3 without a gauge field in the direct formulation,
and the reason for this will become clear in the next subsection.

\subsubsection{Superfluid}

Notice from (\ref{dlxy}) that $\mu$ induces a negative mass term $-\mu^2 |\phi|^2 $. 
So the most likely consequence is that we obtain a superfluid phase with a $\phi$ condensate.

In the $S$ dual formulation, we see that $\mu$ induces a net magnetic flux $\langle \epsilon_{ij} \partial_i a_j \rangle$.
Assuming there is no Higgs phase with a $\psi$ condensate, one consequence is that the spectrum of $\psi$ quanta has the form of gapped Landau levels. And so the superfluid
phase is characterized by
\beq
\langle \phi \rangle \neq 0 \quad, \quad \langle \psi \rangle  = 0. \label{sf}
\eeq

The broken U(1) symmetry due to the $\phi$ condensate implies that there is a gapless Goldstone boson. 
In the $S$ dual formulation, this gapless mode is the $a_\mu$ photon.

\subsubsection{Solid}
\label{sec:solid}

If quantum fluctuations are sufficiently strong, it is possible that for certain CFTs a solid phase obtained.
For the $XY$ model, note that $\mu$ lowers the energy of the $\phi$ particle, while raising
the energy of its anti-particle; so we can consider a low-energy theory of the particle alone.
One possible non-zero density ground state is a crystal of these particles. Clearly such a phase preserves
the global U(1) symmetry, while breaking translational symmetry. 

In the $S$ dual formulation, the solid is obtained by treating $\mathcal{L}_{XY}^S[\mu]$ in the classical
limit, and allowing for a $\psi$ condensate in a Higgs phase. Indeed, this Lagrangian is precisely that for an Abrikosov flux lattice
in the Landau-Ginzburg theory \cite{leefisher}. So we obtain a spatially modulated solution for $\langle \psi \rangle$ in the form
of a triangular lattice. Abrikosov's argument also determines the size of the unit cell of this lattice. To keep the argument general, let us imagine
that the field $\phi$ has electric charge $\mathcal{Q} = q_e$ (the present model has $q_e = 1$), and that the field $\psi$ has magnetic charge 
$\widetilde{\mathcal{Q}} = q_m$ (present model has $q_m = 2 \pi$), and the total area of the system is $L^2$. 
The average flux density $\langle \epsilon_{ij} \partial_i a_j \rangle$ equals $q_e \langle \mathcal{Q} \rangle /(2 \pi L^2)$ via the $S$ duality
relation (\ref{defjmu2}), and so Abrikosov's condition of a flux quantum per unit cell implies
\beq
q_e q_m \frac{\langle \mathcal{Q} \rangle}{L^2} A  = 2 \pi,
\eeq
where $A$ is the area of the unit cell. This corresponds to a U(1) charge $\mathcal{Q} = 1$ per unit cell.

So we see that the solid phase is characterized by
\beq
\langle \phi \rangle = 0 \quad, \quad \langle \psi \rangle  \neq 0. \label{solid}
\eeq
and, as claimed, the monopole operator $\psi$ is the solid order parameter.

\subsection{Abelian $\mathbb{CP}^1$ model}
\label{sec:dcp}

We will apply a chemical potential, $\mu$, to the $\mathcal{Q}_1$ charge only. Then
\bea
\mathcal{L}_{\mathbb{CP}} [\mu] &=& [(\partial_\tau + i b_\tau + \mu) z_1^\ast][ (\partial_\tau - i b_\tau - \mu)z_1 ] + 
[(\partial_\tau + i b_\tau - \mu) z_2^\ast][ (\partial_\tau - i b_\tau + \mu)z_2 ] \nn 
&~&+ \left|\left(\partial_i - i b_i \right) z_1 \right|^2 +
\left|\left(\partial_i - i b_i \right) z_2 \right|^2  + \frac{1}{2e^2} \left( \epsilon_{\mu\nu\lambda} \partial_\nu b_\lambda \right)^2
+ \mathcal{L}_{z,{\rm loc}} 
\eea
and in the $S$ dual theory
\beq
\mathcal{L}_{\mathbb{CP}}^S [\mu] = \mathcal{L}_{\mathbb{CP}}^S - \frac{\mu}{\pi} \epsilon_{ij} \partial_i a_j
\eeq

The superfluid and solid phases of $\mathcal{L}_{\mathbb{CP}}[\mu]$ have a structure similar to that of the $XY$ model,
and so our discussion of these will be brief. 

The superfluid phase has a condensate of $z_1$ and $z_2$, and hence a gauge-invariant
condensate of $\phi_1$. The $b_\mu$ gauge field is Higgsed. There is a gapless Goldstone mode associated with the broken 
$\mathcal{Q}_1$ symmetry, and this is $S$ dual to the $a_\mu$ photon. And the $w_1$ and $w_2$ quanta are gapped. 

As in the $XY$ model,  if quantum fluctuations are sufficiently strong, it is possible that a solid phase obtained.
For the abelian $\mathbb{CP}^1$ model, note that $\mu$ lowers the energy of the $z_1$ particles  and the $z_2$ anti-particles, while raising
the energy of $z_1$ anti-particles  and the $z_2$ particles; so we can consider a low-energy theory of the $z_1$ ($z_2$) particles (anti-particles) alone.
These excitations carry opposite charges under the $b_\mu$ gauge field. So one possible ground state
with a non-zero $\mathcal{Q}_1$ density is a crystalline arrangement of these charges.
Note that while the $z_1$ and $z_2$ excitations carry opposite $b_\mu$ charges, they carry the same $\mathcal{Q}_1$ charge, and this
prevents them from annihilating each other. In the $S$ dual formulation, the solid appears as a Abrikosov flux lattice,
but in a theory with 2 ``superconducting'' order parameters \cite{babaev}; there is flux $\langle \epsilon_{ij} \partial_i a_j \rangle$
of $\pi$ per unit cell, and this corresponds to a charge of $\mathcal{Q}_1 = 2$, one each for the $z_1$ particles and $z_2$ anti-particles.

\subsubsection{Bose metal}
\label{sec:bose}

However, the most interesting feature is the possibility of a compressible phase which is neither a solid nor a superfluid, and which breaks no symmetries.
As we noted above, $\mu$ prefers particles of $z_1$ and anti-particles $z_2$, and so let us write $\mathcal{L}_{\mathbb{CP}}[\mu]$ in a non-relativistic
approximation by integrating out the anti-particles of $z_1$ and the particles of $z_2$:
\bea
\mathcal{L}_{\mathbb{CP}}^{\rm nr} &=& z_1^\ast \left( \partial_\tau - i b_\tau - \mu \right) z_1 + z_2 \left( \partial_\tau + i b_\tau - \mu \right) z_2^\ast \nn
&~&
- \frac{1}{2m}  z_1^\ast \left(\partial_i - i b_i \right)^2 z_1  - \frac{1}{2m}
z_2 \left(\partial_i + i b_i \right)^2 z_2^\ast + \ldots \nn
&~& + \frac{1}{2e^2} \left( \epsilon_{\mu\nu\lambda} \partial_\nu b_\lambda \right)^2
+ \mathcal{L}_{z,{\rm loc}} , \label{lnr}
\eea
where the ellipses represent higher order terms in the bosons kinetic energy, and we expect by scaling that the boson effective mass $m \sim \mu$.
Now we can apply an {\em exact\/} transformation which fermionizes the $z_1$ and $z_2$ quanta by attaching $2\pi$ gauge flux tube of another U(1) gauge field $c_\mu$ \cite{shapere,wilczek}.
We write this transformation as
\bea
f_1 &=& \mathcal{F}_c \, z_1 \nn
f_2 &=& \mathcal{F}_c \, z_2^\ast \label{fluxtube}
\eea
where $\mathcal{F}_c$ is flux tube attachment operator \cite{maissam}. Note the formal analogy to the 
relativistic monopole flux 
operator $\mathcal{M}_b$ which was attached to $z_{1,2}$ in (\ref{dyons}). In the present non-relativistic context,
we attach a flux tube, and this converts non-relativistic bosons to non-relativistic fermions. 
The Lagrangian for the fermions is
\bea
\mathcal{L}_{\mathbb{CP}}^{\rm nr} &=& f_1^\dagger \left( \partial_\tau - i b_\tau - i c_\tau - \mu \right) f_1 
+ f_2^\dagger \left( \partial_\tau + i b_\tau + i c_\tau - \mu \right) f_2 \nn
&~&- \frac{1}{2m}  f_1^\dagger \left(\partial_i - i b_i - i c_i \right)^2 f_1  - \frac{1}{2m}
f_2^\dagger \left(\partial_i + i b_i + i c_i \right)^2 f_2 + \ldots \nn
 \nn
&~& + \frac{i}{4 \pi} \epsilon_{\mu\nu\lambda} c_\mu \partial_\nu c_\lambda + \frac{1}{2e^2} \left( \epsilon_{\mu\nu\lambda} \partial_\nu b_\lambda \right)^2
+ \mathcal{L}_{f,{\rm loc}}, \label{Lfermions}
\eea
and (\ref{lnr}) and (\ref{Lfermions}) are exactly equivalent.
Note the Chern-Simons term in the $c_\mu$ gauge field.

A key feature of (\ref{Lfermions}) is the equation of motion of $c_\tau$:
\beq
f_1^\dagger f_1 - f_2^\dagger f_2 = \frac{1}{2\pi} \epsilon_{ij} \partial_i c_j
\eeq
This implies that in a state with $\langle f_1^\dagger f_1 \rangle = \langle f_2^\dagger f_2 \rangle$, which corresponds to the compressible
phases we are interested in, the net $c_\mu$ flux will be zero, and the $f_1$ and $f_2$ fermions move in a net zero magnetic field.
This is a key feature which allows the Bose metal phase here. And it is this step that fails when we apply the fermionization transformation to the $XY$ model.

To proceed, we follow \cite{hermele}: we map $b_\mu \rightarrow b_\mu - c_\mu$, and then integrate out the Gaussian $c_\mu$ fluctuations. 
Then, dropping higher derivative terms, we obtain a theory without a Chern-Simons term
\bea
\mathcal{L}_{\mathbb{CP}}^{\rm nr} &=& f_1^\dagger \left( \partial_\tau - i b_\tau - \mu \right) f_1 
+ f_2^\dagger \left( \partial_\tau + i b_\tau  - \mu \right) f_2 \nn
&~&- \frac{1}{2m}  f_1^\dagger \left(\partial_i - i b_i \right)^2 f_1  - \frac{1}{2m}
f_2^\dagger \left(\partial_i + i b_i  \right)^2 f_2 + \ldots + \mathcal{L}_{f,{\rm loc}} \label{Lfermions2}
\eea
This describes a compressible Bose metal \cite{motfish}. Of course, there is the possibility that there is a pairing instability with a
$f_1 f_2$ condensate. This will lead to a superfluid state, but this is not identical to the superfluid discussed earlier in this section; the present 
superfluid has a gapless $b_\mu$ photon mode, which was not present earlier.
We assume that this superfluid instability is somehow suppressed to a low energy scale.
In \cite{motfish} this is accomplished by endowing the $f_1$ and $f_2$ fermions with different Fermi surface shapes, and some analog
of this may be possible here.

\section{Holography}
\label{sec:ads}

We begin with a discussion of step B in Fig.~\ref{fig:flowchart}.

First, let us consider the holographic representation of the $XY$ CFT3 on AdS$_4$. 
We propose that the bulk theory should have fields corresponding to 
the conserved current, and to the electric and magnetic operators:
\beq
J_\mu \rightarrow A_a \quad, \quad \phi \rightarrow \Phi \quad , \quad \psi \,\mathcal{M}_a \rightarrow \Psi . \label{JA}
\eeq
Also we note Witten's observation \cite{witten} that $S$ duality on the boundary theory corresponds to
electromagnetic duality in the bulk theory. This suggests that $\psi \, \mathcal{M}_a$ couples to the electromagnetic
dual of $A_a$, which we denote $\widetilde{A}_a$: so we have the following minimal structure of the action
\beq
\mathcal{S}_{XY} = \int d^4 x \sqrt{g} \left[ \frac{1}{4g^2} F_{ab}^2 + |(\partial_a - i A_a) \Phi|^2 + m_e^2 |\Phi|^2 
+ |(\partial_a - i 2 \pi  \widetilde{A}_a) \Psi|^2 + m_m^2 |\Psi|^2 
\right] \label{SXY}
\eeq
where
\beq
F_{ab} = \partial_a A_b - \partial_b A_a \quad , \quad \widetilde{F}_{ab} = \partial_a \widetilde{A}_b - \partial_b \widetilde{A}_a 
\quad , \quad \widetilde{F}_{ab} = \frac{i}{2} \epsilon_{abcd} F^{cd}. \label{FA}
\eeq
The factor of $i$ is needed in the last expression as in (\ref{defjmu2}); it is also
connected to $S^2 = -1$ \cite{witten}, and will be important in Appendix~\ref{app:3pt}.
The mass $m_e$ is determined by the scaling dimension of the electric operator, the mass $m_m$ by the scaling dimension
of the magnetic operator, and $g^2$ is the universal number in (\ref{JJ}). We have not written out the gravitational sector of the action,
along with other possible neutral scalars. Indeed (\ref{SXY}) should be considered a minimal theory with bulk fields which 
correspond to the simplest primary operators of the $XY$ model, and so describes models similar to the $XY$ model. We are not attempting to obtain the full holographic equivalent of the $XY$ CFT3.

Note that there is a non-local relationship between the vector potentials $A_a$ and $\widetilde{A}_a$ in (\ref{FA}).
Here $A_a$ is related to the local observable $J_\mu$ in (\ref{JA}). So, clearly, the non-locality of $\widetilde{A}_a$ 
is linked to the non-locality in the definition of the monopole operator $\psi \, \mathcal{M}_a$ which was noted in Section~\ref{sec:xy}.
An important check of the coupling between $\widetilde{A}_a$ and the monopole field $\Psi$ in (\ref{SXY})
is provided by its predictions for the 3-point correlator of $A_a$
with $\Psi$. We compute this in Appendix~\ref{app:3pt}, and show that the holographic result
has the same form as the corresponding CFT3 correlator of $J_\mu$ and $\psi \, \mathcal{M}_a$. 

For the abelian $\mathbb{CP}^1$ model, the analogous proposal has 
two copies of this structure, and the boundary $\rightarrow$ bulk correspondence
leads to dyonic operators
\bea
&& J_{1\mu} \rightarrow A_a \quad, \quad z_1 \, \mathcal{M}_b \rightarrow Z_1 \quad , \quad z_2 \, \mathcal{M}_b \rightarrow Z_2 \nn
&& J_{2\mu} \rightarrow B_a \quad, \quad w_1 \, \mathcal{M}_a \rightarrow W_1 \quad , \quad w_2 \, \mathcal{M}_a \rightarrow W_2 ,
\eea
with the minimal action
\bea
\mathcal{S}_{\mathbb{CP}} &=& \int d^4 x \sqrt{g} \Biggl[ \frac{1}{4g^2} F_{ab}^2 +\frac{1}{4g^2} G_{ab}^2 + 
|(\partial_a - i A_a -  i\pi  \widetilde{B}_a) Z_1|^2 + m^2 |Z_1|^2 \nn &+& 
|(\partial_a + i A_a - i \pi  \widetilde{B}_a) Z_2|^2 + m^2 |Z_2|^2 +
|(\partial_a - i B_a -  i\pi  \widetilde{A}_a) W_1|^2 + m^2 |W_1|^2 \nn &+& 
|(\partial_a + i B_a - i \pi  \widetilde{A}_a) W_2|^2 + m^2 |W_2|^2 
\Biggr] \label{SCP}
\eea
where as in (\ref{FA})
\beq
G_{ab} = \partial_a B_b - \partial_b B_a \quad , \quad \widetilde{G}_{ab} = \partial_a \widetilde{B}_b - \partial_b \widetilde{B}_a 
\quad , \quad \widetilde{G}_{ab} = \frac{i}{2} \epsilon_{abcd} G^{cd}. \label{GB}
\eeq
Note that to this order, there is 
no direct coupling between the $\mathcal{Q}_1$ and $\mathcal{Q}_2$ sectors
in $\mathcal{S}_{\mathbb{CP}}$ apart from their common coupling to gravitation. The simplest terms are prohibited by the $\mathbb{Z}_2$ symmetry
under which $z_1 \leftrightarrow z_2$, which was mentioned earlier. Dyonic operators also appeared \cite{rey,burgess} in holographic
studies of the quantum Hall effect, but it that case they carried electric and magnetic charges of the same gauge field;
in our case, the dyons carry electric charges under one gauge field, and magnetic charges of a second gauge field.

We are now ready to turn to step C of Fig.~\ref{fig:flowchart}.

The generalization to the non-zero $\mu$ case is now immediate. We simply apply the chemical potential as a boundary condition to
$A_\tau$ \cite{myers}. We can also add various dilaton fields and potentials, as appropriate for the IR metric \cite{kiritsis,trivedi,tadashi1,hyper}.

Let us now discuss the possible phases of $\mathcal{S}_{XY}$. A state with superfluid order has a $\Phi$ condensate \cite{gubser}.
With the monopole $\Psi$ in hand, here we can also obtain a phase with crystalline order, in a manner similar to that for the
boundary theory in Section~\ref{sec:solid}. Notice that with an applied chemical potential, there is an electric field in $\langle F_{ab} \rangle$,
which translates into a magnetic field in $\langle \widetilde{F}_{ab} \rangle$. So the  bulk theory for $\Psi$ is the same as that of an electrically
charged scalar moving in a background magnetic field. The condensation of $\Psi$ in such a situation has been considered
in supergravity theories \cite{jerome3}, and leads to a vortex lattice \cite{jerome3,vortex},
the bulk analog of the Abrikosov flux lattice. In terms of the original direct variables, this clearly corresponds to a boundary
state with crystalline order. However, as we noted in Section~\ref{sec:intro}, these superfluid and solid phases are, strictly speaking,
{\em not\/} the conventional superfluids or solids of Section~\ref{sec:dxy}.
They contain a horizon in the infrared, and so correspond
to `fractionalized' phases with additional deconfined excitations.

Indeed, it is best to think of the symmetry-broken phases above as descending from, and retaining many of the features of, the symmetric
phase with no condensate or broken symmetry. So let us turn to a characterization of such a possible symmetric state
in which none of the fields $\Phi, \Psi, Z_1, Z_2, W_1, W_2$ condense, as may be arranged by making their masses very large.
The most natural conclusion from our 
previous analysis of the boundary theory for the abelian $\mathbb{CP}^1$ model is that such a symmetric phase is a Bose metal, or related non-Fermi liquid. It is useful to define the Bose metal in gauge-invariant terms, to help identify it in the holographic theory:
the Bose metal is a compressible phase with gapless excitations at all momenta, accompanied by signatures of a gauge-charged Fermi surface, which
include ({\em i\/}) Friedel oscillations in the density correlations at the extremal wavevector, $2 k_F$, of the gauge-charged Fermi surface;
({\em ii\/}) logarithmic violation of the area law of the entanglement entropy, with a co-efficient fixed by the charge density \cite{tadashi1,hyper}.
A strong form of the conjecture of \cite{liza,hyper,solvay} is that all compressible phases which do not break any symmetry
are ultimately Bose metals, Fermi liquids, or allied phases with visible and/or gauge-charged Fermi surfaces.

So can theories like those in (\ref{SXY}, \ref{SCP}) describe Bose metals? We leave the answer of this question to future work,
and just make some general remarks here.
As discussed earlier \cite{tadashi1,hyper}, the entanglement entropy for certain hyperscaling violating backgrounds has numerous features
consistent with a gauge-charged Fermi surface. So a key question is whether the holographic theories (\ref{SXY}, \ref{SCP}), or their extensions,
contain Friedel oscillations.
It is clear that information on the oscillatory structure is already present: after all, condensates of the monopole fields $\Psi, W_1, W_2$, 
leads to crystalline order with precisely the right
period of an integer number of particles per unit cell. So we need to make $\Psi, W_1, W_2$ ``almost'' condensed to obtain Friedel oscillations.
Furthermore, the theories  (\ref{SXY}, \ref{SCP}) are similar to theories of vortex liquids in classical superconductors in an applied
magnetic field: the latter systems have been studied using Feynman graph expansions \cite{brezin}, density-functional theories \cite{tesanovic,menon} and numerical simulations \cite{ceperly}, and show clear oscillatory
structure in the vortex-vortex correlation functions. From these studies, we can expect that the bulk
$\widetilde{F}_{ab}$ correlations will have a structure factor with a maximum at a non-zero wavevector, and the boundary limit of this structure factor
is the density-density correlator of the doped CFT3. 

\section{Conclusions}
\label{sec:conc}

We have presented an analysis of possible phases of doped CFT3s. The $S$ duality properties of the parent CFT3, and its electric and magnetic
operators were important in our analysis, and for our proposed bulk theory on AdS$_4$.
We found that the doped CFT3s had phases with superfluid and solid order, and a Bose metal phase which broke
no symmetries. The magnetic operators of the parent CFT3 served as order parameters for the solid, and also determined the 
size of its unit cell.

We checked the structure of the bulk theory on AdS$_4$ by both bulk and boundary 
computations of 3-point correlators in Appendix~\ref{app:3pt}. 
We exhibited connections between the Bose metal phases of doped CFT3s and the holographic compressible 
phase with no broken symmetries on asymptotically AdS$_4$ spacetimes, and these were summarized
in Section~\ref{sec:ads}. The magnetic operators of the CFT3 translated
into new terms in the holographic theory which are sensitive to the quantization of particle number, and produce associated periodic
correlations in the density.
 We also noted that holographic states with broken symmetries are best understood as
Bose metals upon which a broken symmetry has been superimposed: thus holographic superfluids and solids have broken particle number
and translational symmetries respectively, concomitant with the excitations of a deconfined gauge theory.
 
The key step in obtaining a Bose metal in our doped CFT3 was the flux {\em tube} 
attachment operator 
$\mathcal{F}_c$ \cite{maissam} in (\ref{fluxtube}) which converted 
 the non-relativistic bosons $z_1$ and $z_2^\ast$ in (\ref{lnr}) to non-relativistic fermions $f_1$ and $f_2$ in (\ref{Lfermions}). This operation has a 
formal similarity to a relativistic analog in our discussion of the CFT3 in (\ref{dyons}),
where the {\em monopole} flux operator $\mathcal{M}_b$ (defined in (\ref{ma})) 
was attached to the relativistic fields $z_1$ and $z_2$ to obtain gauge-invariant
primary fields of the CFT3. The relativistic $\mathcal{M}_b$ operator has proposed counterparts in 
bulk monopole/dyon fields
on AdS$_4$, as discussed in Section~\ref{sec:ads}.
We now need to understand the holographic extension of $\mathcal{F}_c$ better, by finding additional
signatures of the gauge-charged fermions in the Bose metal.

\subsection{Generalizations}
\label{sec:gen}

The explicit analyses of this paper have been for CFT3s with a global U(1) symmetry which are also Abelian
gauge theories, and are expressible using only bosonic degrees of freedom. We conclude by briefly noting 
the applicability of our results to other CFT3s with a global U(1) symmetry. 

The analyses defining the monopole operator in direct representation of the $XY$ model in Section~\ref{sec:xy} generalize
to any CFT3 with a global U(1). We can always gauge the global U(1) as in (\ref{lxyalpha}), and insert monopole sources
in the background gauge field. Consequently, we expect there to be an analog of the bulk field $\Psi$ in (\ref{SXY}) for all
such CFT3s. The global U(1) current $J_\mu$ will be holographically dual to a bulk U(1) gauge field $A_a$, and $\Psi$ will
be electrically coupled to the $S$ dual gauge field $\widetilde{A}_a$, just as in (\ref{SXY}). It is $\Psi$ that carries the information
on periodic density modulations in the compressible phases, and so these are ubiquitous, as expected.
However, the field $\Phi$ in (\ref{SXY})
is not as ubiquitous: its existence requires the presence of a gauge-invariant CFT3 operator carrying the global U(1) charge, and these need not be 
present, or could carry large enough dimensions to be irrelevant. 

When we restrict attentions to CFT3s with a global U(1) which are also Abelian gauge theories, then further applications of our result are possible, even
in cases where explicit $S$ dual
mappings are not known. As an example, we can consider a CFT3 with Dirac fermions, such as those in \cite{liza}, 
obtained by replacing $z_1, z_2$ in $\mathcal{L}_{\mathbb{CP}}$ (Eq. (\ref{lcp})) by two-component Dirac fermions, $q_1, q_2$.
In this case also, as in (\ref{dyons}), we will have gauge-invariant operators $q_1 \, \mathcal{M}_b$, $q_2 \, \mathcal{M}_b$
carrying electric charge $\mathcal{Q}_1$ and magnetic charge $\widetilde{\mathcal{Q}}_2$, even though their $S$ dual counterparts
are not evident. And there should be fermionic bulk operators, $Q_1, Q_2$ with the same quantum numbers which could 
reveal the ``hidden'' Fermi surfaces of Bose metal-like phases in these Abelian gauge theories. 
However, this construction of gauge-invariant operators carrying the fundamental
global electric charge does not appear to generalize to non-Abelian gauge theories.

\acknowledgements

I thank T.~Faulkner, M.~P.~A.~Fisher, J.~Gauntlett, S.~Hartnoll, L.~Huijse, J.~Hung, N.~Iqbal, S.~Kachru, 
I.~Klebanov, Sung-Sik Lee, D.~R.~Nelson,  
B.~Swingle, X.~Yin, and especially A.~Kapustin and S.~Raju for valuable discussions. 
This research was supported by the National Science Foundation under grant DMR-1103860, and by 
the Army Research Office Award W911NF-12-1-0227.

\appendix

\section{$S$ duality of the $XY\/$ model}
\label{app:xy}

We review the duality mapping of the $XY$ model \cite{peskin,dasgupta} in (\ref{lxyalpha}). We begin by writing $\phi \sim e^{i \theta}$,
and expressing the action in the Villain form on a cubic lattice of sites $i$:
\beq
\mathcal{L}_{XY} [\alpha] = \frac{K}{2} \left( \Delta_\mu \theta_i - \alpha_{i \mu} - 2 \pi n_{i \mu} \right)^2, \label{xyvillain}
\eeq
where $\Delta_\mu$ is a discrete lattice derivative, $n_{i \mu}$ are integers on the links of the cubic lattice,
and $\alpha_{i \mu}$ is the monopole background field. We perform a Fourier transform and write
\beq
\mathcal{L}_{XY} [\alpha]  = \frac{1}{2K} J_{i \mu}^2 +  i J_{i \mu} \left( \Delta_\mu \theta_i - \alpha_{i \mu} \right) \label{axy3}
\eeq
where $J_{i\mu}$ are another set of integers on the links of the cubic lattice. Integrating over $\theta_i$, we obtain the zero
divergence condition $\Delta_\mu J_{i \mu} = 0$. We solve this condition by writing 
\beq
J_{i \mu} = \frac{1}{2\pi} \epsilon_{\mu\nu\lambda} \Delta_\nu a_{\jmath\lambda}
\eeq
where $a_{\jmath \mu}$ resides on the links of the dual cubic lattice with sites $\jmath$, and takes values which are integer multiples of $2 \pi$.
Then we have
\beq
\mathcal{L}_{XY} [\alpha] = \frac{1}{8 \pi^2 K} \left( \epsilon_{\mu\nu\lambda} \Delta_\nu a_{\jmath\lambda} \right)^2 
- \frac{i}{2 \pi}  a_{\jmath\mu} \beta_{\jmath\mu}, \label{axy2}
\eeq
where $\beta_{\jmath\mu} = \epsilon_{\mu\nu\lambda} \Delta_\nu \alpha_\lambda$ is the magnetic flux associated with the monopole insertion.
At this point, we can drop the ``Dirac string'' 
contribution to $\beta_{\jmath\mu}$ because it only changes the Lagrangian by integer multiples of $2 \pi i$.

So far, everything has been an exact rewriting of (\ref{xyvillain}). Now we promote $a_{\jmath\mu}$ to a continuous real field by writing
\beq
\mathcal{L}_{XY}^S [\alpha] = \frac{1}{8 \pi^2 K} \left( \epsilon_{\mu\nu\lambda} \Delta_\nu a_{\jmath\lambda} \right)^2 
- y \cos (a_{\jmath \mu} )
- \frac{i}{2 \pi}  a_{\jmath\mu} \beta_{\jmath\mu}. \label{axy1}
\eeq
Note that (\ref{axy1}) is exactly equivalent to (\ref{axy2}) in the limit $y \rightarrow \infty$. But, as argued in \cite{dasgupta}, 
the physics at finite $y$ is the same
as that as $y \rightarrow \infty$, and so will work with the $S$ dual Lagrangian $\mathcal{L}_{XY}[\alpha]$. We can make the Lagrangian
have the structure of a gauge theory by the shift $a_{\jmath\mu} \rightarrow a_{\jmath\mu} - \Delta_\mu \vartheta_\jmath$, where $\vartheta_\jmath$ is a dual angular variable. The integral over $\vartheta_\jmath$ only introduces a redundancy on configuration space, and the original
expression merely corresponds to the gauge choice $\vartheta_{j} = 0$. Finally, taking the continuum limit with $\psi \sim e^{i \vartheta}$,
we obtain the $S$ dual Lagrangian (\ref{xyS}), and also the correlator (\ref{mxyb}) after using (\ref{divbeta}).

Conversely, let us begin with a lattice version of the $S$ dual theory in the presence of a monopole background, $\mathcal{L}_{XY}^S [\alpha]$
in (\ref{lsxyalpha}):
\beq
\mathcal{L}_{XY}^S [\alpha] = \frac{y}{2} \left( \Delta_\mu \vartheta_\jmath - a_{\jmath \mu} - \alpha_{\jmath \mu} - 2 \pi n_{\jmath \mu} \right)^2
+ \frac{1}{8 \pi^2 K} \left( \epsilon_{\mu\nu\lambda} \Delta_\nu a_{\jmath\lambda} \right)^2
, \label{xysvillain}
\eeq
After similar steps, this maps exactly to
\beq
\mathcal{L}_{XY}^S [\alpha] = \frac{1}{8 \pi^2 y} \left( \epsilon_{\mu\nu\lambda} \Delta_\nu b_{i\lambda} \right)^2
+ \frac{K}{2} \left( \Delta_\mu \gamma_i - b_{i \mu} \right)^2  - \frac{i}{2\pi} b_{i \mu} \beta_{i \mu}
, \label{xysvillain2}
\eeq
where $b_{i \mu}$ takes values which are integer multiples of $2 \pi$ on the links of the direct lattice, and $\gamma_i$ is a real
variable on the direct lattice. Then promoting $b_{i \mu}$ to a real variable, and shifting $b_{i \mu} \rightarrow b_{i \mu} - \Delta_\mu \theta_i$,
$\gamma_i \rightarrow \gamma_i - \theta_i$,
as below (\ref{axy2}), we obtain
\beq
\mathcal{L}_{XY} [\alpha] = \frac{1}{8 \pi^2 y} \left( \epsilon_{\mu\nu\lambda} \Delta_\nu b_{i\lambda} \right)^2
+ \frac{K}{2} \left( \Delta_\mu \gamma_i - b_{i \mu} \right)^2  - \frac{i}{2\pi} (b_{i \mu} - \Delta_\mu \theta_i) \beta_{i \mu} - y \cos (\Delta_\mu \theta_i - b_{i \mu} ).
\label{xysvillain3}
\eeq
This expression shows that the $b_\mu$ gauge field has been Higgsed by $e^{i \gamma}$, and so ignoring the massive Higgs mode
we can set $b_{i \mu} = 0$ in the gauge $\gamma_i = 0$. The resulting theory is just a lattice version of the XY model of (\ref{lxy})
with $\phi \sim e^{i \theta}$. Upon using (\ref{divbeta}),  
the effect of the monopole insertion $\alpha_{\jmath\mu}$ is to yield the electric correlator in (\ref{exy}).

Other relationships between the correlators of the direct and $S$ dual theories can be obtained in a similar manner, by inserting appropriate
sources in the starting Lagrangian.

\section{Three point correlators of the $XY$ model}
\label{app:3pt}

\begin{figure}[h]
\begin{center}
 \includegraphics[width=4in]{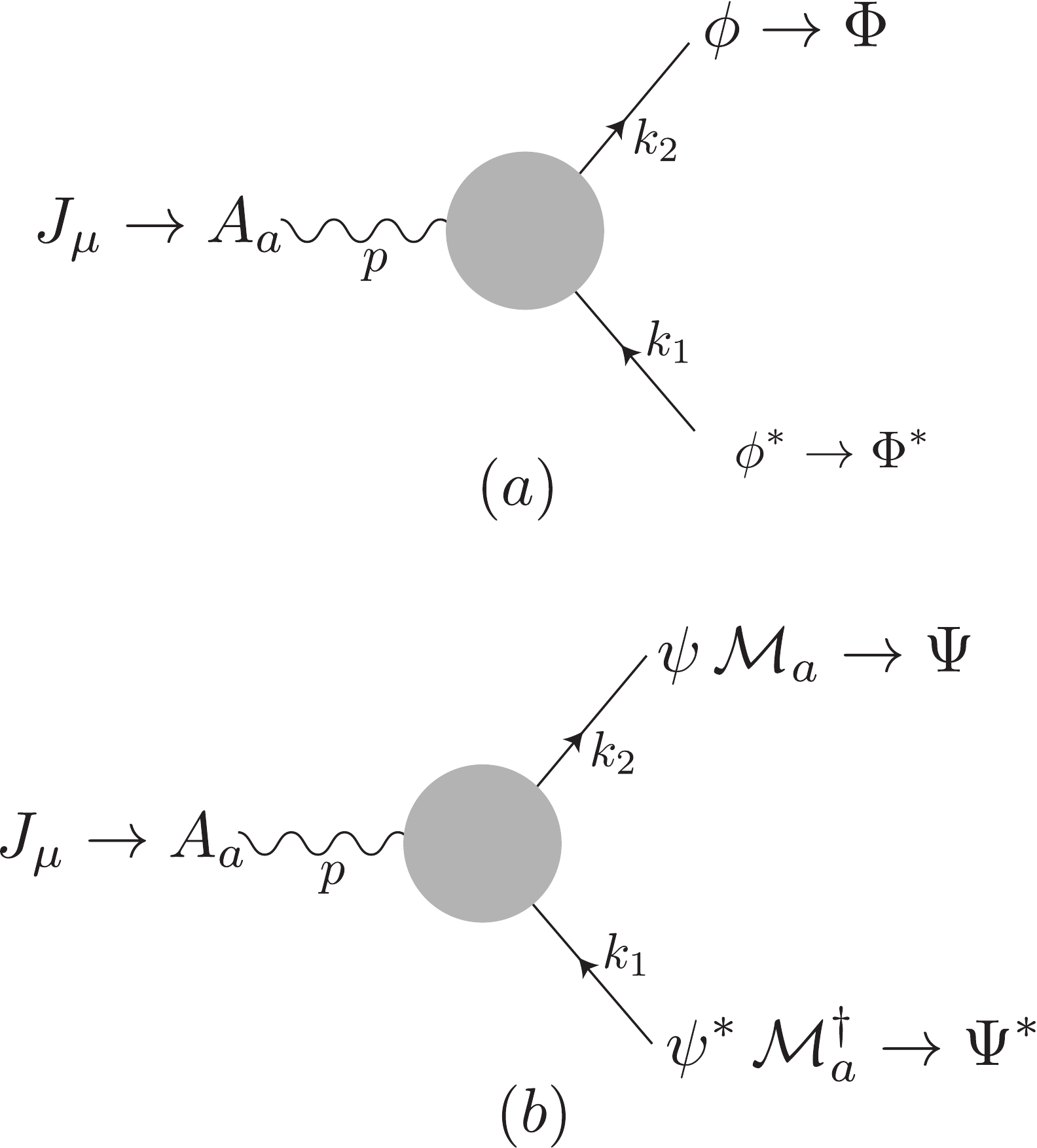}
 \caption{3-point correlators of the $XY$ model: ({\em a}\/) the electric correlator $K$ in (\ref{K1}),
 ({\em b}\/) the magnetic correlator $K_m$ in (\ref{K4}). The labels are the boundary $\rightarrow$ bulk fields.}
\label{fig:3pt}
\end{center}
\end{figure}

First, we compute the 3-point correlator between the conserved current $J_\mu$ and the electrically charged field $\phi$
of the $XY$ model in (\ref{xy3a}) shown in Fig.~\ref{fig:3pt}. 
For the free CFT3, or in the large flavor number limit of the interacting CFT3, this is
\bea
K (p, k_1, k_2) &=& \varepsilon_{\mu} (p) \left\langle J_\mu (p) \phi^\ast (k_1) \phi (k_2) \right\rangle_{\mathcal{L}_{XY}} \nn
&=& \frac{\varepsilon_{\mu} (p) (k_{1\mu} + k_{2\mu})}{k_1^2 k_2^2}  \label{K1}
\eea
where $p_\mu \equiv k_{1\mu} - k_{2 \mu}$, and $\varepsilon_{\mu} (p)$ is a polarization vector orthogonal to $p_\mu$,
$\varepsilon_{\mu} (p) p_\mu = 0$.

Let us now compare the result (\ref{K1}) with that obtained by a
 tree-level holographic computation from the bulk 
action $\mathcal{S}_{XY}$ in (\ref{SXY}). We will label the holographic direction $z$ and the AdS$_4$ metric
\beq
ds^2 = \frac{dz^2 + dx_\mu^2}{z^2}.
\eeq
The correlator is given by the 3-point interaction in the action $\mathcal{S}_{XY}$, evaluated with
the bulk fields taking values specified by the boundary-bulk propagators \cite{Wittenads,kw,rastelli,suvrat1}; 
in the gauge $A_z = 0$, $\partial_\mu A_\mu = 0$ the bulk fields are
\bea
\Phi (k, z) &=&
|k|^{\Delta_e - \frac{3}{2}} z^{3/2} K_{\Delta_e - \frac{3}{2}} ( |k| z ) \nn
A_\mu (p,z) &=& \varepsilon_{\mu} (p) e^{- |p| z}, \label{K2} 
\eea
where $\Delta_e$ is the scaling dimension of the $\phi$ field.
Then the 3-point correlator is 
\bea 
K (p, k_1, k_2) &=& \int_0^{\infty} \frac{dz}{z^2}  (k_{1 \mu} + k_{2 \mu} ) \Phi^\ast (k_1, z) \Phi (k_2, z)  A_\mu (p , z)  \nn
&=&  \frac{\varepsilon_{\mu} (p) (k_{1\mu} + k_{2\mu})}{|k_1|^{5/2-\Delta_e} |k_2|^{5/2-\Delta_e}} 
\mathcal{F} \left( \frac{|k_1|}{|p|}, \frac{|k_2|}{|p|} \right) \label{K3}
\eea
where $\mathcal{F}$ is a dimensionless function of its dimensionless arguments whose value can be deduced from the expressions above.
Notice the similarity between the vector structure of the expressions in (\ref{K1}) and (\ref{K3}).
It is expected that the two results will match when the CFT3 computation is extended beyond the free field limit to non-trivial values of $\Delta_e$.

We now extend these computations to the 3-point correlator between the current $J_\mu$ and the monopole operator of the $XY$ model,
specified in (\ref{xy3b}) and shown in Fig.~\ref{fig:3pt}.
On the CFT3 side, it is difficult to work with (\ref{xy3b}), and 
this computation is more easily performed using the $S$ dual representation of Section~\ref{sec:sxy}. Under this mapping, using the
transformations of Appendix~\ref{app:xy},  (\ref{xy3b})
becomes the 
gauge-invariant correlator 
\bea
K_m (p,k_1,k_2) &=& \int d^3 y d^3 w \, e^{i p w - i k_1 y} \varepsilon_{\mu} (p) \nn
&~&~~~~~~~\times
\left\langle J_\mu (w)  \psi^\ast (y) \exp \left( -\frac{i}{2 \pi} \int d^3 x \, a_\mu \, \beta_\mu \right) \psi (0) \right\rangle_{\mathcal{L}_{XY}^S} , \label{K4}
\eea
where the current $J_\mu$ is now given by (\ref{defjmu2}). This correlator is best computed in the large $N_f$ limit of the CFT3s in which $\psi$ has $N_f$ flavors. 
Then in the leading large $N_f$ limit the transverse gauge field propagator is \cite{kauln}
\beq
\left\langle a_\mu (p) a_\nu (p) \right\rangle = \frac{16}{N_f |p|} \left( \delta_{\mu\nu} - \frac{p_\mu p_\nu}{p^2} \right) . \label{K5}
\eeq
Evaluating (\ref{K4}) to order $1/N_f$ with this propagator we obtain
\beq
K_m (p, k_1, k_2) = \left(\frac{16}{N_f \pi} \right) \frac{ \epsilon_{\mu\nu\lambda} \varepsilon_{\mu} (p) k_{1\nu} k_{2 \lambda}}{k_1^2 k_2^2 |p|} . \label{K6}
\eeq
We note that the exponential factor in (\ref{K4}) has a vanishing contribution at this order, and (\ref{K6}) arises only from the current vertex of $\psi$.

Finally, let us compare (\ref{K6}) with the tree-level holographic computation from the bulk 
action $\mathcal{S}_{XY}$ in (\ref{SXY}). Just as in (\ref{K2}), we will now have the bulk fields
\bea
\Psi (k, z) &=& 
|k|^{\Delta_m - \frac{3}{2}} z^{3/2} K_{\Delta_m - \frac{3}{2}} ( |k| z ) \nn
A_\mu (p,z) &=& \varepsilon_{\mu} (p) e^{- |p| z}, \label{K7} 
\eea
where $\Delta_m$ is the scaling dimension of $\psi \, \mathcal{M}_a$.
We now need to convert the above result for $A_\mu$ in (\ref{K7}) to an expression for $\widetilde{A}_{\mu}$.
Let the holographic indices $a,b \ldots$ extend over the directions $z,x_1,x_2,x_3$, and let us choose
the momentum $p_\mu = (p,0,0)$ (with $p>0$) and $\varepsilon_\mu = (0,1,0)$. 
Then, the Maxwell tensor is
\beq
F_{12} = i p e^{-p z} \quad, \quad F_{z2} = - p e^{-p z}
\eeq
and all other components are zero. So, from (\ref{FA}), we have the dual tensor
\beq
\widetilde{F}_{3z} = - p e^{-p z} \quad, \quad \widetilde{F}_{13} = - i p e^{-p z},
\eeq
which corresponds to a dual vector potential
\beq
\widetilde{A}_\mu (p, z) = - (0,0,1) e^{-p z}. 
\eeq
From this we deduce the following result for general $p_\mu$ and $\varepsilon_\mu$:
\beq
\widetilde{A}_\mu (p,z) =  \frac{\epsilon_{\mu\nu\lambda} p_\nu \varepsilon_\lambda (p) }{|p|} e^{-|p| z}, \label{K10}
\eeq
and $\widetilde{A}_z = 0$.
Note that the simple exponential form of the boundary-bulk correlator of the gauge field was crucial in the above analysis leading to
the simple result for the dual gauge field in (\ref{K10}).
We can now obtain the 3-point correlator as in (\ref{K3})
\bea 
K_m (p, k_1, k_2) &=& \int_0^{\infty} \frac{dz}{z^2}  (k_{1 \mu} + k_{2 \mu} ) \Psi^\ast (k_1, z) \Psi (k_2, z)  \widetilde{A}_\mu (p , z)  \nn
&=&  \frac{\epsilon_{\mu\nu\lambda} \varepsilon_{\mu} (p) k_{1\nu} k_{2 \lambda}}{|k_1|^{5/2-\Delta_m} |k_2|^{5/2-\Delta_m}|p|} 
\widetilde{\mathcal{F}} \left( \frac{|k_1|}{|p|}, \frac{|k_2|}{|p|} \right), \label{K8}
\eea
where again $\widetilde{\mathcal{F}}$ is a dimensionless function of its dimensionless arguments whose value can be deduced from the expressions above.
Now notice the remarkable match of the magnetic holographic result (\ref{K8}) to the CFT3 computation in (\ref{K6}), similar to that for the electric operator case between (\ref{K1}) and (\ref{K3}).

\section{$S$ duality of the abelian $\mathbb{CP}^1$ model}
\label{app:cp1}

We proceed just as in Appendix~\ref{app:xy}, following \cite{mv}. We start from the $S$ dual action 
(\ref{lcps}), write $w_{1,2} \sim e^{i \vartheta_{1,2}}$, and introduce the Villain action on the dual cubic lattice
\bea
\mathcal{L}_{\mathbb{CP}}^S [\alpha] &=& \frac{K}{2} \left( \Delta_\mu \vartheta_{1\jmath} - a_{\jmath\mu} - \eta_1 \alpha_{\jmath \mu} - 2 \pi n_{1\jmath\mu} \right)^2 + \frac{K}{2} \left( \Delta_\mu \vartheta_{2\jmath} - a_{\jmath\mu} - \eta_2 \alpha_{\jmath \mu} - 2 \pi n_{2\jmath\mu} \right)^2 \nn
&~& + \frac{1}{2e^2} \left( \epsilon_{\mu\nu\lambda} \partial_\nu a_{\jmath\lambda} \right)^2 .
\label{cpvillain}
\eea
Here $\alpha_{\jmath \mu}$ is a monopole background field defined by (\ref{betaalpha}) and (\ref{divbeta}).
The choices of $\eta_{1,2} = 0, \pm 1$ will give expressions for the different operator insertions. Thus, the choice $\eta_1 = 1$, $\eta_2 = 0$
yields (\ref{lcp1}).

We begin with a Fourier transform, as in (\ref{axy3}), to obtain
\bea
\mathcal{L}_{\mathbb{CP}}^S [\alpha] &=& \frac{1}{2K} \left( J_{1\jmath \mu}^2 + J_{2\jmath\mu}^2 \right)  - i a_{\jmath \mu} \left( J_{1\jmath\mu} + J_{2\jmath \mu} \right)
- i \alpha_{\jmath \mu} \left( \eta_1 J_{1\jmath\mu} + \eta_2 J_{2\jmath\mu} \right)  \nn
&~& + \frac{e^2}{2} f_{i \mu}^2 + i f_{i \mu} \epsilon_{\mu\nu\lambda} \Delta_\nu a_{\jmath\lambda} 
\eea
where $J_{1i\mu}$ and $J_{2i\mu}$ are integer valued currents obeying $\Delta_\mu J_{1i\mu} = \Delta_\mu J_{2 i \mu} = 0$,
and $f_{\jmath \mu}$ is a real-valued flux on the links of the dual lattice. Integrating over $a_{\jmath \mu}$ we obtain the additional constraint
$J_{1i \mu} + J_{2 i \mu} = \epsilon_{\mu\nu\lambda} \Delta_\nu f_{\jmath \lambda}$.
We solve these constraints by writing
\bea
J_{1\jmath \mu} &=& \frac{1}{2\pi} \epsilon_{\mu\nu\lambda} \Delta_\nu b_{1i\lambda} \nn
J_{2\jmath \mu} &=& \frac{1}{2\pi} \epsilon_{\mu\nu\lambda} \Delta_\nu b_{2i\lambda} \nn
f_{\jmath\mu} &=& \frac{1}{2\pi} \left( b_{1i\mu} + b_{2i\mu} - \Delta_\mu \gamma_i \right)
\eea
where $b_{1i\mu}$ and $b_{2i \mu}$ are integer multiples of $ 2 \pi$, and $\gamma_i$ is real-valued. So the action is 
\bea
\mathcal{L}_{\mathbb{CP}}^S [\alpha] &=& \frac{1}{8\pi^2K} \left( \left( \epsilon_{\mu\nu\lambda} \Delta_\nu b_{1i\lambda} \right)^2 + \left( \epsilon_{\mu\nu\lambda} \Delta_\nu b_{2i\lambda} \right)^2 \right)  
-  \frac{i}{2\pi}    \beta_{i \mu} \left(\eta_1 b_{1i\mu} + \eta_2 b_{2i\mu} \right)\nn
&~& + \frac{e^2}{8 \pi^2} \left( b_{1i\mu} + b_{2i\mu} - \Delta_\mu \gamma_i \right)^2 
\eea
As in Appendix~\ref{app:xy}, we can now drop the Dirac string in $\beta_{i\mu}$ because it only changes
the action by integer multiples of $2 \pi i$. Also, 
up to this point, all transformations have been exact. 

Now we promote $b_{1i\mu}$ and $b_{2i\mu}$
to continuous real fields, and shift $b_{1i\mu} \rightarrow b_{1i\mu} - \Delta_\mu \theta_{1j}$,
$b_{2i\mu} \rightarrow b_{2i\mu} - \Delta_\mu \theta_{2j}$, and $\gamma_i \rightarrow \gamma_i - \theta_{1i} - \theta_{2 i}$. Then we obtain the direct lattice theory
\bea
 \mathcal{L}_{\mathbb{CP}} [\alpha] &=& \frac{1}{8\pi^2K} \left( \left( \epsilon_{\mu\nu\lambda} \Delta_\nu b_{1i\lambda} \right)^2 + \left( \epsilon_{\mu\nu\lambda} \Delta_\nu b_{2i\lambda} \right)^2 \right)  \nn
&-&  \frac{i}{2\pi} \beta_{i \mu} \left[ \eta_1 \left(b_{1i\mu}  - \Delta_\mu \theta_{1i}  \right) + \eta_2 \left(b_{2i\mu}  - \Delta_\mu \theta_{2i}  \right) \right]  \nn
&+& \frac{e^2}{8 \pi^2} \left( b_{1i\mu} + b_{2i\mu} - \Delta_\mu \gamma_i \right)^2 - y \left(
\cos (\Delta_\mu \theta_{1 i \mu} - b_{1 i \mu}) + \cos (\Delta_\mu \theta_{2 i \mu} - b_{2 i \mu}) \right).
\label{betasign}
\eea
This action has the structure of a U(1)$\times$U(1) gauge theory, in the presence of charged 
matter fields, $e^{i \theta_1}$ and $e^{i \theta_2}$. One of the diagonal U(1)s has been Higgsed by the term proportional to $e^2$,
with $e^{i \gamma}$ acting as the Higgs field. So we can drop the massive excitations associated with this diagonal U(1) by setting
$b_{1i\mu} = - b_{2 i \mu} = b_{i \mu}$ in the gauge $\gamma_i = 0$. Then, in the continuum limit
with $z_1 \sim e^{i \theta_1}$ and $z_2 \sim e^{- i \theta_2}$, and $\eta_1=\eta_2 = 0$, 
we obtain the action of the abelian $\mathbb{CP}^1$ model
in (\ref{lcp}). Other values of $\eta_1 , \eta_2$ can now be used to establish the duality mappings of the operator insertions, and we note
typical examples
\begin{itemize}
\item $\eta_1=1$, $\eta_2=0$: This establishes the equality between the correlators in (\ref{z1a}) and (\ref{z1b}) upon applying (\ref{divbeta}).
\item $\eta_1 = 1$, $\eta_2 = -1$: This monopole flux couples to the global $\mathcal{Q}_2$ charge of (\ref{lcps}), and so corresponds to operator insertions 
with $\widetilde{\mathcal{Q}}_2 = 2 \pi$, and all other electric and magnetic charges equal to zero. The above analysis shows that 
this is the two-point correlator of $z_1 z_2 \, \mathcal{M}_b^2$, 
corresponding to the operator identification in (\ref{psimab}).
\item $\eta_1 = 1$, $\eta_2 = 1$: This is a monopole gauge flux in the $a_\mu$ gauge field, and so via (\ref{J1a}) only carries $\mathcal{Q}_1=2$ electrical charge.
Above we find the correlator $z_2^\ast z_1$, which carries the expected charge, as in (\ref{phi12}). 
\end{itemize}


\begin{thebibliography}{}

\bibitem{nernst}
  S.~A.~Hartnoll, P.~K.~Kovtun, M.~M\"uller, and S.~Sachdev,
  ``Theory of the Nernst effect near quantum phase transitions in condensed matter, and in dyonic black holes,''
  Phys.\ Rev.\  {\bf B76}, 144502 (2007)
  [arXiv:0706.3215 [cond-mat.str-el]].

\bibitem{Maldacena}
  J.~M.~Maldacena,
  ``The Large $N$ limit of superconformal field theories and supergravity,''
  Adv.\ Theor.\ Math.\ Phys.\  {\bf 2}, 231 (1998)
  [Int.\ J.\ Theor.\ Phys.\  {\bf 38}, 1113 (1999)]
  [arXiv:hep-th/9711200].

\bibitem{GKP}
  S.~S.~Gubser, I.~R.~Klebanov and A.~M.~Polyakov,
  ``Gauge theory correlators from noncritical string theory,''
  Phys.\ Lett.\ B {\bf 428}, 105 (1998)
  [arXiv:hep-th/9802109].

\bibitem{Wittenads}
  E.~Witten,
  ``Anti-de Sitter space and holography,''
  Adv.\ Theor.\ Math.\ Phys.\  {\bf 2}, 253 (1998)
  [arXiv:hep-th/9802150].
  
\bibitem{gubser}   S.~S.~Gubser,
  ``Breaking an abelian gauge symmetry near a black hole horizon,''
  Phys.\ Rev.\ D {\bf 78}, 065034 (2008)
  [arXiv:0801.2977 [hep-th]].

\bibitem{myers}
  A.~Chamblin, R.~Emparan, C.~V.~Johnson and R.~C.~Myers,
  ``Charged AdS black holes and catastrophic holography,''
  Phys.\ Rev.\ D {\bf 60}, 064018 (1999)
  [arXiv:hep-th/9902170].

\bibitem{tadashi}   T.~Nishioka, S.~Ryu, and T.~Takayanagi,
  ``Holographic Superconductor/Insulator Transition at Zero Temperature,''
  JHEP {\bf 1003}, 131 (2010)
  [arXiv:0911.0962 [hep-th]].

\bibitem{gary} G.~T.~Horowitz and B.~Way,
  ``Complete Phase Diagrams for a Holographic Superconductor/Insulator System,''
  JHEP {\bf 1011}, 011 (2010)
  [arXiv:1007.3714 [hep-th]].

\bibitem{liza} L.~Huijse and S.~Sachdev, ``Fermi surfaces and gauge-gravity duality,''
Phys. Rev. D {\bf 84}, 026001 (2011) [arXiv:1104.5022 [hep-th]].

\bibitem{hyper}  L. Huijse, S. Sachdev, and B. Swingle, 
``Hidden Fermi surfaces in compressible states of gauge-gravity duality,''
Phys. Rev. B {\bf 85}, 035121 (2012) [arXiv:1112.0573 [cond-mat.str-el]].

\bibitem{tomjoe}   T.~Faulkner and J.~Polchinski,
  ``Semi-Holographic Fermi Liquids,''
  JHEP {\bf 1106}, 012 (2011)
  [arXiv:1001.5049 [hep-th]].

\bibitem{ssffl}   S.~Sachdev,
  ``Holographic metals and the fractionalized Fermi liquid,''
  Phys.\ Rev.\ Lett.\  {\bf 105}, 151602 (2010)
  [arXiv:1006.3794 [hep-th]].

\bibitem{ssfl} S.~Sachdev,   ``A model of a Fermi liquid using gauge-gravity duality,''
  Phys.\ Rev.\  D {\bf 84}, 066009 (2011)
  [arXiv:1107.5321 [hep-th]].

\bibitem{arcmp} S.~Sachdev,  ``What can gauge-gravity duality teach us about condensed matter physics?''
Annual Review of Condensed Matter Physics
{\bf 3}, 9 (2012)  [arXiv:1108.1197 [cond-mat.str-el]].

\bibitem{hartnollrev}   S.~A.~Hartnoll,
  ``Horizons, holography and condensed matter,''
  arXiv:1106.4324 [hep-th].

\bibitem{iqballutt}   N.~Iqbal and H.~Liu,
  ``Luttinger's Theorem, Superfluid Vortices, and Holography,''
  arXiv:1112.3671 [hep-th].

\bibitem{lizasean}    S.~A.~Hartnoll and L.~Huijse,
  ``Fractionalization of holographic Fermi surfaces,''
  arXiv:1111.2606 [hep-th].

\bibitem{mcgreevy}   A.~Allais, J.~McGreevy and S.~J.~Suh,
  ``A quantum electron star,''
  arXiv:1202.5308 [hep-th].

\bibitem{seanrad}   S.~A.~Hartnoll and D.~Radicevic,
  ``Holographic order parameter for charge fractionalization,''
  arXiv:1205.5291 [hep-th].

\bibitem{kiritsis}   C.~Charmousis, B.~Gouteraux, B.~S.~Kim, E.~Kiritsis and R.~Meyer,
  ``Effective Holographic Theories for low-temperature condensed matter systems,''
  JHEP {\bf 1011}, 151 (2010)
  [arXiv:1005.4690 [hep-th]].

\bibitem{trivedi}   N.~Iizuka, N.~Kundu, P.~Narayan and S.~P.~Trivedi,
  ``Holographic Fermi and Non-Fermi Liquids with Transitions in Dilaton Gravity,''
  JHEP {\bf 1201}, 094 (2012)
  [arXiv:1105.1162 [hep-th]].

\bibitem{tadashi1}   N.~Ogawa, T.~Takayanagi and T.~Ugajin,
  ``Holographic Fermi Surfaces and Entanglement Entropy,''
  JHEP {\bf 1201}, 125 (2012).
  [arXiv:1111.1023 [hep-th]]

\bibitem{ooguri1}   S.~Nakamura, H.~Ooguri and C.~-S.~Park,
  ``Gravity Dual of Spatially Modulated Phase,''
  Phys.\ Rev.\ D {\bf 81}, 044018 (2010)
  [arXiv:0911.0679 [hep-th]].

\bibitem{ooguri2}   H.~Ooguri and C.~-S.~Park,
  ``Holographic End-Point of Spatially Modulated Phase Transition,''
  Phys.\ Rev.\ D {\bf 82}, 126001 (2010)
  [arXiv:1007.3737 [hep-th]].

\bibitem{jerome1}   A.~Donos and J.~P.~Gauntlett,
  ``Holographic striped phases,''
  JHEP {\bf 1108}, 140 (2011)
  [arXiv:1106.2004 [hep-th]].

\bibitem{jerome2}   A.~Donos, J.~P.~Gauntlett and C.~Pantelidou,
  ``Spatially modulated instabilities of magnetic black branes,''
  JHEP {\bf 1201}, 061 (2012)
  [arXiv:1109.0471 [hep-th]].

\bibitem{jerome3}   A.~Donos, J.~P.~Gauntlett and C.~Pantelidou,
  ``Magnetic and Electric AdS Solutions in String- and M-Theory,''
  arXiv:1112.4195 [hep-th].

\bibitem{iizuka}   N.~Iizuka, S.~Kachru, N.~Kundu, P.~Narayan, N.~Sircar and S.~P.~Trivedi,
  ``Bianchi Attractors: A Classification of Extremal Black Brane Geometries,''
  JHEP {\bf 1207}, 193 (2012)
  [arXiv:1201.4861 [hep-th]].

\bibitem{witten} E.~Witten,
  ``SL(2,Z) action on three-dimensional conformal field theories with abelian
  symmetry,''
  arXiv:hep-th/0307041.

\bibitem{m2cft}   C.~P.~Herzog, P.~Kovtun, S.~Sachdev and D.~T.~Son,
  ``Quantum critical transport, duality, and M-theory,''
  Phys.\ Rev.\ D {\bf 75}, 085020 (2007)
  [arXiv:hep-th/0701036].

\bibitem{fi}   T.~Faulkner and N.~Iqbal,
  ``Friedel oscillations and horizon charge in 1D holographic liquids,''
  arXiv:1207.4208 [hep-th].

\bibitem{rajesh}   R.~Gopakumar, A.~Hashimoto, I.~R.~Klebanov, S.~Sachdev and K.~Schoutens,
  ``Strange Metals in One Spatial Dimension,''
  arXiv:1206.4719 [hep-th].

\bibitem{murthy} G.~Murthy and S.~Sachdev, Nucl. Phys. B {\bf 344}, 557
(1990).

\bibitem{kapustin1}   A.~Kapustin and M.~J.~Strassler,
  ``On mirror symmetry in three-dimensional abelian gauge theories,''
  JHEP {\bf 9904}, 021 (1999)
  [arXiv:hep-th/9902033].

\bibitem{kapustin2}   V.~Borokhov, A.~Kapustin and X.-k.~Wu,
  ``Topological disorder operators in three-dimensional conformal field theory,''
  JHEP {\bf 0211}, 049 (2002)
  [arXiv:hep-th/0206054].

\bibitem{kapustin3}   V.~Borokhov, A.~Kapustin and X.-k.~Wu,
  ``Monopole operators and mirror symmetry in three-dimensions,''
  JHEP {\bf 0212}, 044 (2002)
  [arXiv:hep-th/0207074].

\bibitem{hermelemono} M.~A.~Metlitski, M.~Hermele, T.~Senthil, 
and M.~P.~A.~Fisher, ``Monopoles in $\mathbb{CP}^{N-1}$ model via the state-operator correspondence,''
Phys. Rev. B {\bf 78}, 214418 (2008) [arXiv:0809.2816 [cond-mat.str-el]].

\bibitem{hermeleo4}  M.~Hermele, 
``Non-abelian descendant of abelian duality in a two-dimensional frustrated quantum magnet,''
Phys. Rev. B {\bf 79}, 184429 (2009)
[arXiv:0902.1350 [cond-mat.str-el]].

\bibitem{benna}   M.~K.~Benna, I.~R.~Klebanov and T.~Klose,
  ``Charges of Monopole Operators in Chern-Simons Yang-Mills Theory,''
  JHEP {\bf 1001}, 110 (2010)
  [arXiv:0906.3008 [hep-th]].

\bibitem{willett}   A.~Kapustin and B.~Willett,
  ``Generalized Superconformal Index for Three Dimensional Field Theories,''
  arXiv:1106.2484 [hep-th].

\bibitem{rsl} N.~Read and S.~Sachdev, ``Valence bond and spin-Peierls ground states of low dimensional quantum antiferromagnets,''
Phys. Rev. Lett. {\bf 62}, 1694 (1989).

\bibitem{rsb} N. Read and S. Sachdev, ``Spin-Peierls, valence bond solid, and Neel ground states of low dimensional quantum antiferromagnets,''  Phys. Rev. B {\bf 42}, 4568 (1990). 

\bibitem{senthil} T.~Senthil, A.~Vishwanath, L.~Balents, S.~Sachdev, and
M.~P.~A.~Fisher, ``Deconfined quantum critical points,'' Science
{\bf 303}, 1490 (2004) [arXiv:cond-mat/0311326] ; T.~Senthil,
L.~Balents, S.~Sachdev, A.~Vishwanath, and M.~P.~A.~Fisher,
``Quantum criticality beyond the Landau-Ginzburg-Wilson paradigm,''
Phys.\ Rev.\ B {\bf 70}, 144407 (2004) [arXiv:cond-mat/0312617].

\bibitem{leefisher} M.~P.~A.~Fisher and D.~H.~Lee, 
``Correspondence between two-dimensional bosons and a bulk superconductor in a magnetic field,''
Phys. Rev. B {\bf 39}, 2756 (1989).

\bibitem{motfish} O.~I~ Motrunich and M.~P.~A.~Fisher, ``D-wave correlated Critical Bose Liquids in two dimensions,''
Phys. Rev. B {\bf 75}, 235116 (2007) [arXiv:cond-mat/0703261].

\bibitem{motfish2} M.~S.~Block, R.~V.~Mishmash, R.~K.~Kaul, D.~N.~Sheng, O.~I.~Motrunich, and M.~P.~A.~Fisher,
``Exotic Gapless Mott Insulators of Bosons on Multi-Leg Ladders,'' 
Phys. Rev. Lett. {\bf 106}, 046402 (2011) [arXiv:1008.4105 [cond-mat.str-el]].

\bibitem{motfish3} M.~S.~Block, D.~N.~Sheng, O.~I.~Motrunich, and M.~P.~A.~Fisher, ``Spin Bose-Metal and Valence Bond 
Solid phases in a spin-1/2 model with ring exchanges on a four-leg triangular ladder,'' 
Phys. Rev. Lett. {\bf 106}, 157202 (2011) [arXiv:1009.1179 [cond-mat.str-el]].

\bibitem{motfish4} R.~V.~Mishmash, M.~S.~Block, R.~K.~Kaul, D.~N.~Sheng, O.~I.~Motrunich, 
and M.~P.~A.~Fisher, ``Bose Metals and Insulators on Multi-Leg Ladders with Ring Exchange,''
Phys. Rev. B {\bf 84}, 245127 (2011)
 [arXiv:1110.4607 [cond-mat.str-el]].
 
\bibitem{peskin} M.~E.~Peskin, ``Mandelstam-'t Hooft duality in abelian lattice
models,'' Annals of Physics, {\bf 113}, 122 (1978).

\bibitem{dasgupta} C.~Dasgupta and B.~I.~Halperin, ``Phase Transition in a
Lattice Model of Superconductivity,'' Phys.\ Rev.\ Lett.\ {\bf 47},
1556 (1981).

\bibitem{dirac} P.~A.~M.~Dirac, ``Gauge-invariant formulation of quantum electrodynamics,'' Can. J. Phys. {\bf 33}, 650 (1955).

\bibitem{sondhi} T.~H.~Hansson, V.~Oganesyan, and S.~L.~Sondhi, ``Superconductors are topologically ordered,'', 
Annals of Physics {\bf 313}, 497 (2004) [arXiv:cond-mat/0404327 [cond-mat.supr-con]].

\bibitem{rantner}
W.~Rantner and X.-G.~Wen, ``Gauge invariance and electron spectral functions in underdoped cuprates,'' 
arXiv:cond-mat/0105540

\bibitem{jinwu} J.~Ye, ``On gauge-invariant Green function in 2+1 dimensional QED,''
Phys. Rev. B {\bf 67}, 115104 (2003) [arXiv:cond-mat/0205417].

\bibitem{mpaf2} M.~P.~A.~Fisher, ``Quantum phase transitions in disordered
two-dimensional superconductors,'' Phys.\ Rev.\ Lett.\ {\bf 65}, 923
(1990).

\bibitem{wenzee} X.-G.~Wen and A.~Zee, ``Universal conductance at the
superconductor-insulator transition,'' Int. J. Mod. Phys. B {\bf 4},
437 (1990).

\bibitem{max1} M. A. Metlitski and S. Sachdev, ``Valence bond solid order near impurities in two-dimensional quantum antiferromagnets,''  
Phys. Rev. B {\bf 77}, 054411 (2008) [arXiv:0710.0626 [cond-mat.str-el]].
      
\bibitem{mv} O.~I.~Motrunich and A.~Vishwanath, ``Emergent photons and new
transitions in the O(3) sigma model with hedgehog suppression,''
Phys.\ Rev.\ B {\bf 70}, 075104 (2004) [arXiv:cond-mat/0311222].

\bibitem{balents} L. Balents, L. Bartosch, A. Burkov, S. Sachdev, and K. Sengupta, 
``Putting competing orders in their place near the Mott transition, ,''
Phy. Rev. B {\bf 71}, 144508 (2005) [arXiv:cond-mat/0408329].

\bibitem{mv2} O.~I.~Motrunich and A.~Vishwanath, 
``Comparative study of Higgs transition in one-component and two-component lattice superconductor models,''
arXiv:0805.1494 [cond-mat.str-el] and references therein.

\bibitem{babaev}  J.~Carlstrom, E.~Babaev, and M.~Speight, 
``Type-1.5 superconductivity in multiband systems: the effects of interband couplings,''
Phys. Rev. B {\bf 83}, 174509 (2011)
[arXiv:1009.2196 [cond-mat.supr-con]].

\bibitem{shapere} {\em Geometric Phases in Physics},  A.~Shapere and F.~Wilczek, Eds.,
Advanced Series in Mathematical Physics, Vol 5, World Scientific, Singapore (1988).

\bibitem{wilczek} {\em Fractional Statistics and Anyon Superconductivity\/}, F. Wilczek ed., 
World Scientific (1990).

\bibitem{maissam} M.~Barkeshli and J.~McGreevy,
``Continuous transitions between composite Fermi liquid and Landau Fermi liquid: a route to fractionalized Mott insulators,''
Phys. Rev. B {\bf 86}, 075136 (2012) [arXiv:1206.6530 [cond-mat.str-el]].

\bibitem{hermele} J.~Alicea, O.~I.~Motrunich, M.~Hermele, and M.~P.~A.~Fisher, ``Criticality in quantum triangular antiferromagnets via fermionized vortices,'' Phys. Rev. B {\bf 72}, 064407 (2005) [arXiv:cond-mat/0503399].

\bibitem{rey}   D.~Bak and S.~-J.~Rey,
  ``Composite Fermion Metals from Dyon Black Holes and S-Duality,''
  JHEP {\bf 1009}, 032 (2010)
  [arXiv:0912.0939 [hep-th]].

\bibitem{burgess}  A.~Bayntun, C.~P.~Burgess, B.~P.~Dolan and S.~-S.~Lee,
  ``AdS/QHE: Towards a Holographic Description of Quantum Hall Experiments,''
  New J.\ Phys.\  {\bf 13}, 035012 (2011)
  [arXiv:1008.1917 [hep-th]].

\bibitem{vortex}   O.~Domenech, M.~Montull, A.~Pomarol, A.~Salvio and P.~J.~Silva,
  ``Emergent Gauge Fields in Holographic Superconductors,''
  JHEP {\bf 1008}, 033 (2010)
  [arXiv:1005.1776 [hep-th]].

\bibitem{solvay}  S. Sachdev, ``The quantum phases of matter,''  25th Solvay Conference on Physics, "The Theory of the Quantum World", Brussels, Oct 2011, arXiv:1203.4565 [hep-th].

\bibitem{brezin} E.~Br\'ezin, D.~R.~Nelson, and A.~Thiaville, ``Fluctuation effects near $H_{c2}$ in type-II superconductors,'' 
Phys. Rev. B {\bf 31}, 7124 (1985).

\bibitem{tesanovic} I.~F.~Herbut and Z.~Te\v{s}anovi\'{c}, ``Density-Functional Theory of Freezing of Vortex Liquid in Quasi-Two-Dimensional Superconductors,'' Phys. Rev. Lett. {\bf 73}, 484 (1994).

\bibitem{menon} G.~I.~Menon, C.~Dasgupta, H.~R.~Krishnamurthy, T.~V.~Ramakrishnan, and S.~Sengupta, 
``Density-functional theory of flux-lattice melting in high-$T_c$ superconductors,'' Phys. Rev. B {\bf 54}, 16192 (1996).

\bibitem{ceperly} W. R. Magro and 
D. M. Ceperley, ``Ground state of two-dimensional Yukawa bosons: Applications to vortex melting,'' Phys. Rev. B {\bf 48}, 411 (1993).


\bibitem{kw}   I.~R.~Klebanov and E.~Witten,
  ``AdS / CFT correspondence and symmetry breaking,''
  Nucl.\ Phys.\ B {\bf 556}, 89 (1999)
  [arXiv:hep-th/9905104].

\bibitem{rastelli}   T.~Hartman and L.~Rastelli,
  ``Double-trace deformations, mixed boundary conditions and functional determinants in AdS/CFT,''
  JHEP {\bf 0801}, 019 (2008)
  [arXiv:hep-th/0602106].

\bibitem{suvrat1}   S.~Raju,
  ``Four Point Functions of the Stress Tensor and Conserved Currents in AdS$_4$/CFT$_3$,''
  Phys.\ Rev.\ D {\bf 85}, 126008 (2012)
  [arXiv:1201.6452 [hep-th]].


\bibitem{kauln} R.~K.~Kaul and S.~Sachdev,  
``Quantum criticality of U(1) gauge theories with fermionic and bosonic matter in two spatial dimensions,''
Phys. Rev. B {\bf 77}, 155105 (2008) [arXiv:0801.0723 [cond-mat.str-el]].

\end{thebibliography}
\end{document}